\documentclass[final,5p,twocolumn]{elsarticle}
\usepackage{amssymb}
\usepackage{amsmath}
\usepackage{cuted}
\usepackage{array}
\usepackage{graphicx}
\usepackage{caption}
\usepackage{subcaption}
\usepackage{color}
\usepackage[cjk]{kotex}

\journal{arXiv}

\begin{document}

\begin{frontmatter}



%\fontsize{0.5cm}{0.1cm}\selectfont
\title{Idle speed control with low-complexity offset-free explicit model predictive control in presence of system delay}

\author[a]{Sang Hwan Son}
\author[b]{Se-Kyu Oh}
\author[c]{Byung Jun Park}
\author[c]{Min Jun Song}
\author[c]{Jong Min Lee\corref{cor}}
\address[a]{Artie McFerrin Department of Chemical Engineering, Texas A\&M University, College Station, TX 77845, USA}
\address[b]{Electrification Control Development Team 1, Hyundai Motor Company, Hwaseong-si 18280, South Korea}
\address[c]{School of Chemical and Biological Engineering, Institute of Chemical Processes, Seoul National University, Seoul 08826, South Korea}
\cortext[cor]{Corresponding author's email: jongmin@snu.ac.kr.}

%\author{}

%\address{}

\begin{abstract}
The requirement for continual improvement of idle speed control (ISC) performance is increasing due to the stringent regulation on emission and fuel economy these days. In this regard, a low-complexity offset-free explicit model predictive control (EMPC) with constraint horizon is designed to regulate the idle speed under unmeasured disturbance in presence of system delay with rigorous formulation. Particularly, we developed a high-fidelity 4-stroke gasoline-direct injected spark-ignited engine model based on first-principles and test vehicle driving data, and designed a model predictive ISC system. To handle the delay from intake to torque production, we constructed a control-oriented model with delay augmentation. To reject the influence of torque loss, we implemented the offset-free MPC scheme with disturbance model and estimator. Moreover, to deal with the limited capacity assigned for the controller in the engine control unit and the short sampling instant of the engine system, we formulated a low-complexity multiparametric quadratic program with constraint horizon in presence of system delay in state and input variables, and obtained an explicit solution map. To demonstrate the performance of the designed controller, a series of closed-loop simulations were performed. The developed explicit controller showed proper ISC performance in presence of torque loss and system delay.
\end{abstract}


\end{frontmatter}


\section{Introduction}\label{sec1}

Engine idle speed control (ISC) is a crucial issue in automotive control and continually refined, since it considerably influences the fuel economy, emission, safety, combustion stability, and drivability \cite{39,40,41}. In general, the idle speed is desirable to set as low as possible to reduce the fuel consumption; for instance it is known that the constant volume sample-based fuel economy improves by one mile per gallon as the idle speed decreases by 100 rpm \cite{23}. However, lower idle speed increases noise, vibration, harshness, and possibility of engine stalling \cite{2}. Therefore, it is important to optimize powertrain operations to regulate the idle speed at the set-point with an available quality of combustion and emission, and avoid large engine speed deviation by minimizing the unmeasured disturbance effect to prevent engine stalling \cite{31}. Despite the successful implementation of the idle speed control system in most vehicles, continual improvement of the performance of idle speed control is necessary to meet the increasing stringent regulation on emission and fuel economy under recent eco-friendly policies \cite{26,30,37}.

Engine idle speed control is a complex control problem with multi-objectives, multi-variables and system constraints. In the case of spark ignition (SI) engines, the regulation of idle speed and variables, and torque reserve are achieved by manipulating the air flow rate and spark timing subject to constraints due to the system limits such as combustion stability and engine breathing. Though the simple feedback controllers such as PID loops or pole-placement linear controllers have been widely employed for idle speed regulators, advanced optimal control frameworks, which can deal with the difficulties described above with more sophisticated algorithms, have attracted much attention recently \cite{9,38,42}.

Model predictive control (MPC) is becoming increasingly popular in the field of idle speed control due to its several attractive features \cite{26,4,25}; the specification of objective function is available so the multiple objectives can be considered; multivariable systems can be handled in a systematic way; it allows for the specification of the constraints on system variables, and it can take time-domain constraints into account explicitly; and feedback adjustment is embedded in a receding horizon control manner \cite{29,33,34}. However, owing to the limited computation and memory resources of the engine control unit (ECU), the MPC problem cannot be solved in real-time in the vehicle. To handle this limitation, it is common to directly implement only the explicit solution map of the designed MPC \cite{3,2} by solving multiparametric program off-line as proposed in \cite{15,16,17}.

The main objective of ISC system is to maintain the desired engine idle speed. The main cause of the failure in ISC system is torque loss from various sources such as friction, pumping, air conditioner, steering wheel, automatic transmission, etc. Some of the disturbances are measurable and can be handled by feedforward compensation, but unmeasurable disturbance also always exists. Therefore, ISC systems must take into account the rejection of unmeasured disturbance \cite{4}. Model predictive control systems with offset-free tracking property are designed by augmenting the integration of the error in \cite{3,2,32}. However, since error integration is independent of control algorithm, this method can cause windup in constrained systems even when using MPC \cite{35}. Offset-free MPC with disturbance estimator approach in \cite{10,11,12} is proposed to avoid this problem by augmenting the model with estimated disturbance from an observer. This method does not suffer from windup and has anti-windup effect when the system saturates \cite{36,43,44}. However, there is no rigorous formulation for offset-free explicit model predictive ISC with disturbance estimator approach while considering system delay in state and input variables caused by the delay from intake to torque production. 

To this end, in this work, we designed an offset-free explicit MPC (EMPC) system for ISC using disturbance estimator in presence of system delay. Specifically, first, we developed a high-fidelity mean-value model for engine rotational speed, torque generation, and air mass flow of a 4-stroke SI-gasoline direct injection (SI-GDI) engine based on previous studies \cite{2,9,1} and performed model fitting with the test vehicle driving data. Then, we derived a control-oriented model by linearizing and discretizing the developed engine model. To deal with the system delay, the past state and input variables are also augmented to the model. Based on the obtained control-oriented model, an offset-free MPC system with disturbance model and estimator is designed to drive the engine speed to the desired idle speed set-point while considering the influence of the torque loss. And then, we derived the explicit map from state to solution of the offset-free MPC problem by solving the multiparametric quadratic program (mp-QP) off-line. In this process, we also developed a low-complexity mp-QP formulation with constraint horizon in presence of system delay to decrease the complexity of the explicit solution map for further reduction of the processing power for on-line evaluation and the memory consumption in ECU. The objective of the controller includes the torque reserve with a proper amount of spark efficiency degradation as well as idle speed regulation to complement the actuation delay on air flow control \cite{23,2}. Lastly, we demonstrated the closed-loop ISC performance of the developed explicit controller under the influence of the torque loss and the delay from intake center to torque center.

The rest of this paper is organized as follows. In the next section, the mean value model of the 4-stroke SI-GDI engine is developed based on first-principles. Section~\ref{sec3} presents the parameter estimation of the developed model with test vehicle driving data in various conditions. In Section~\ref{sec4}, the design of low-complexity offset-free EMPC system for ISC in presence of system delay in state and input variables is presented. In Section~\ref{sec5}, closed-loop simulation results and analysis are presented. Lastly, we conclude with a few important remarks in Section~\ref{sec6}.
 
\section{Engine model}\label{sec2}

\begin{figure}[h]
\begin{center}
\includegraphics[width=8cm]{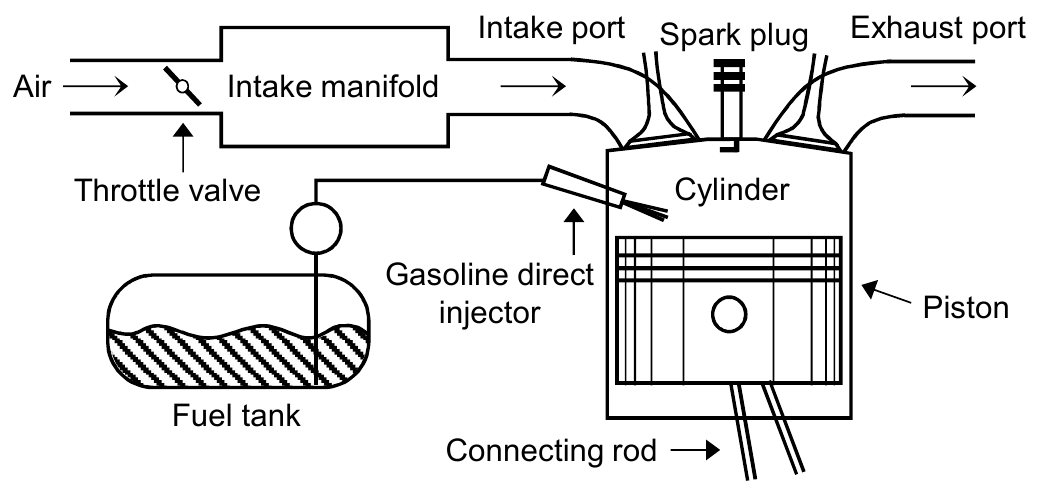}\\
\caption{Schematic illustration of a gasoline direct-injected spark ignition engine.}
\label{fig1}
\end{center}
\end{figure}

Fig.~\ref{fig1} shows the schematic illustration of 4-stroke SI-GDI engine. The air from the air cleaner flows into the intake manifold where the flow rate is controlled by the throttle valve. The air in the intake manifold flows into the cylinder passing through the intake port, then it is mixed and combusted with the injected fuel from the gasoline direct injector where the combustion timing is controlled by the spark plug. As the air-fuel mixture inside the cylinder is combusted, the pressure of combustion gas pushes the piston down and generates mechanical work which is transported to the crankshaft through the connecting rod.

SI-GDI engines are very similar to port-fuel injection SI engines \cite{1}. Therefore, we follow the engine mean-value modeling flow of usual port-fuel injection SI engines which has been widely studied \cite{1,6,7}.

\subsection*{A. Engine rotational speed}
We describe the engine crankshaft rotational dynamics with the Newton's second law:
\begin{align}\label{eq1}
\dot{N}(t)=\frac{60}{\Theta_e}(\varepsilon(t)\cdot M_{m_\varphi}(t)-M_{loss}(t))
\end{align}
where $N$[rpm] is the engine speed, $\dot{N}$[rpm/s] is the engine acceleration, $\Theta_{e}$[kg$\cdot$m$^2$] is the engine inertia, $\varepsilon\in[0,1]$ is the thermodynamic efficiency which denotes the deviation from a perfect conversion, $M_{m_\varphi}$[N$\cdot $m] is the fuel mean effective torque with perfect conversion of the thermal energy of fuel into the mechanical energy, and $M_{loss}$[N$\cdot$m] is the torque loss due to engine drag, electronic accessories, and load on the crankshaft.

\subsection*{B. Torque generation}

The fuel mean effective torque $M_{m_\varphi}$ in (\ref{eq1}) can be described with the fuel mean effective pressure $P_{m_\varphi}$:
\begin{align}\label{eq2}
M_{m_\varphi}(t)=P_{m_\varphi}(t)\frac{V_d}{2}
\end{align}
where $V_d$[m$^3$] is the displaced volume of the cylinder. 

$P_{m_\varphi}$ is the mean effective pressure that engine would generate from the fuel mass with the efficiency of 1:
\begin{align}\label{eq3}
P_{m_\varphi}(t)=\frac{H_\ell\cdot m_\varphi(t)}{V_d}
\end{align}
where $H_\ell$[J/kg] is the lower heating value of the fuel, and $m_\varphi$[kg] is the mass of fuel burnt in a combustion cycle.

$m_\varphi$ can be described with the air mass charged in the cylinder $m_{a,c}$[kg]: 
\begin{align}\label{eq4}
m_\varphi(t)=\frac{m_{a,c}(t)}{\xi\cdot\eta}
\end{align}
where $\xi$ is the stoichiometric air-to-fuel ratio and $\eta$ is the air-to-fuel ratio.

The current cylinder air charge $m_{a,c}(t)$ can be derived from the past air mass flow entering the cylinder and the past engine speed:
\begin{align}\label{eq5}
m_{a,c}(t)=w_{a,c}(t-\tau(t))\frac{2}{N(t-\tau(t))}
\end{align}
where $w_{a,c}$[kg/s] is the cylinder air mass flow, $\tau$[s] is the delay between the air intake and torque production in engine which can be approximated as in \cite{2}:
\begin{align}\label{eq6}
\tau(t)\approx\frac{60}{N(t)}.
\end{align}

Then, substituting (\ref{eq3})--(\ref{eq6}) into (\ref{eq2}) yields:
\begin{align}\label{eq7}
M_{m_\varphi}(t)=\frac{H_\ell\cdot w_{a,c}(t-\tau)}{\xi\cdot\eta\cdot N(t-\tau)}.
\end{align}

The thermodynamic efficiency $\varepsilon$ in (\ref{eq1}) can be separated into each component:
\begin{align}\label{eq8}
\varepsilon(t)=\varepsilon_\zeta(t)\cdot\varepsilon_N(t)
\end{align}
where $\zeta$[$^\circ$] denotes the spark ignition angle. \cite{1} additionally considers the efficiency from the air-to-fuel ratio, compression ratio, exhaust gas recirculation (EGR). However, since the target engine system is designed to have the efficiency values around 1 for air-to-fuel ratio and compression ratio, we do not consider these effects. Additionally, since EGR is not activated in the idle state, we also exclude the influence of EGR.

The spark timing efficiency $\varepsilon_\zeta$ is commonly described as a function of the spark ignition angle deviation from the optimal maximum brake torque (MBT) ignition angle $\zeta^*$ \cite{4,5}:
\begin{align}\label{eq9}
\varepsilon_\zeta(\zeta(t),\zeta^*(t))=cos(\zeta(t)-\zeta^*(t))^\alpha
\end{align}
where $\alpha$ is an engine dependent parameter. Each engine has an intrinsic map of $\zeta^*$ according to the engine operating conditions such as engine speed, cylinder air charge, and load.

The trajectory of the engine speed efficiency $\varepsilon_N$ typically has a parabolic form over the engine speed. The $\varepsilon_N$ reduces at operating condition with very low and high engine speed due to the relatively large heat loss through the wall at low engine speed and the relatively insufficient combustion time from the short interval at high engine speed \cite{1}. In this study, since the control objective is to regulate the idle engine speed, we linearly approximate $\varepsilon_N$ around 700 rpm:
\begin{align}\label{eq10}
\varepsilon_N(N(t))=\beta_0+\beta_1 N(t)
\end{align}
where $\beta_0$ and $\beta_1$ are constant parameters.

\subsection*{C. Air mass flow}

In \cite{1}, the engine air system is regarded as a volumetric pump where the volumetric air flow rate is approximately proportional to the rotational speed of engine. Based on this perspective and ideal gas law, the air mass flow entering the cylinder can be formulated as (\ref{eq11}).
\begin{align}\label{eq11}
w_{a,c}(t)=\frac{P_{im}(t)}{R\cdot T_{im}(t)}\lambda_\nu(t)\cdot V_d\frac{N(t)}{2}
\end{align}
where $P_{im}$[pa] and $T_{im}$[K] are the intake manifold pressure and temperature, $R$ is the ideal gas constant, and $\lambda_\nu$ is the volumetric efficiency. $\lambda_\nu$ denotes the deviation of the engine from the perfect volumetric pump:
\begin{align}\label{eq12}
&\lambda_\nu(t)=\lambda_{\nu,N}(t)\cdot\lambda_{\nu,P}(t)\\
&\lambda_{\nu,N}(t):=\gamma_0+\gamma_1 N(t)+\gamma_2 N(t)^2 \nonumber\\
&\lambda_{\nu,P}(t):=\frac{1}{\eta_c-1}\left( \eta_c-\left(\frac{P_e(t)}{P_{im}(t)}\right)^{1/\delta}\right)\nonumber\\
&\eta_c:=(V_d+V_c)/V_c \nonumber
\end{align}
where $\gamma_0$, $\gamma_1$, $\gamma_2$, and $\delta$ are constant parameters, $P_e$[pa] denotes the exhaust back pressure, $\eta_c$ denotes the compression ratio, and $V_c$[m$^3$] is the compression volume at top dead center.

Intake manifold dynamics is described with air mass flow rate through the electric throttle $w_{a,th}$[kg/s] and $w_{a,c}$ based on the ideal gas law:
\begin{align}\label{eq13}
\dot{P}_{im}(t)=\frac{R\cdot T_{im}}{V_{im}}(w_{a,th}(t)-w_{a,c}(t)).
\end{align}
$w_{a,th}$ can be described with the isentropic expansion \cite{8}:
\begin{align}\label{eq14}
&w_{a,th}(t)=\frac{c_d\cdot P_{a}}{\sqrt{R\cdot T_a}} A_{th}(t)\cdot \mu(t) \\
&\mu(t):=\begin{cases} \sqrt{\kappa(\frac{2}{\kappa+1})^{\frac{\kappa+1}{\kappa-1}}}\quad\mathrm{for}\quad \frac{P_{im}(t)}{P_a} < \left( \frac{2}{\kappa+1} \right) ^{\frac{\kappa}{\kappa-1}} \\ \left( \frac{P_{im}(t)}{P_a} \right)^\frac{1}{\kappa}\sqrt{\frac{2\kappa}{\kappa-1}\left( 1-\left( \frac{P_{im}(t)}{P_a} \right)^\frac{\kappa-1}{\kappa} \right)}\quad\mathrm{otherwise} \end{cases}\nonumber
\end{align}
where $c_d$ is the discharge coefficient, $A_{th}$[m$^2$] is the opening area of throttle which is controlled by the throttle angle $\theta$[$^\circ$], and $P_a$[pa] and $T_a$[K] are the pressure and the temperature of ambient air, respectively. $\mu$ is the flow function \cite{9}, and $\kappa$ is the specific heat ratio.

\section{Parameter estimation}\label{sec3}

We obtained the vehicle driving data with a sampling instant of 0.01 s. The test was conducted under various conditions: no action, window working, steering working, and air conditioner working. Then, we performed model fitting based on the obtained data with the engine model developed in Section~\ref{sec2}.

\subsection*{A. Torque generation}
We derived the spark efficiency $\epsilon_\zeta(t)$ values from the ignition angle measurements. Though the spark efficiency model in (\ref{eq9}) is suggested in \cite{4,5}, we use the given spark efficiency map according to the actual angle difference from the MBT angle of the engine ($\varepsilon_\zeta=f_{\zeta}(\zeta-\zeta^*)$). Then, we derived the delay from intake center to torque center $\tau(t)$ from the engine speed measurement using (\ref{eq6}) and discretized based on the sampling instant.

By combining (\ref{eq7}) and (\ref{eq8}), we describe the engine torque $M_{act}$ as
\begin{align}\label{eq15}
M_{act}(t)&=\varepsilon(t)\cdot M_{m_\varphi}(t) \nonumber\\
&=\varepsilon_\zeta(t)\cdot(\beta_0+\beta_1 N(t))\frac{H_\ell\cdot w_{a,c}(t-\tau(t))}{\xi\cdot\eta\cdot N(t-\tau(t))}.
\end{align}

Then, the parameters $\beta_0$ and $\beta_1$ are estimated by solving the least mean squared error problem with $M_{act},\varepsilon_\zeta,N,w_{a,c}$ and $\tau$ data and given $H_\ell,\xi$ and $\eta$ values:
\begin{align}\label{eq16}
\underset{\small{\beta_0,\beta_1}}{\mathrm{min}}\;  \sum_{i=1}^{n_c} \sum_{j=1}^{n_{t}} ||M_{act}^{(i,j)}-\hat{M}_{act}^{(i,j)}||^2
\end{align}
where $n_c$ is the number of cases, $n_{t}$ is the number of training data set for each case, $i$ and $j$ are the indices for case and sampling instant, respectively, $M_{act}^{(i,j)}$ is $M_{act}$ data of case $i$ at the sampling time $j$, and $\hat{M}_{act}^{(i,j)}$ is the value calculated from (\ref{eq15}) with $\varepsilon_{\zeta}^{(i,j)},N^{(i,j)},w_{a,c}^{(i,j)}$ and $\tau^{(i,j)}$ data. 

\begin{figure}[h]
\begin{center}
\includegraphics[width=9cm]{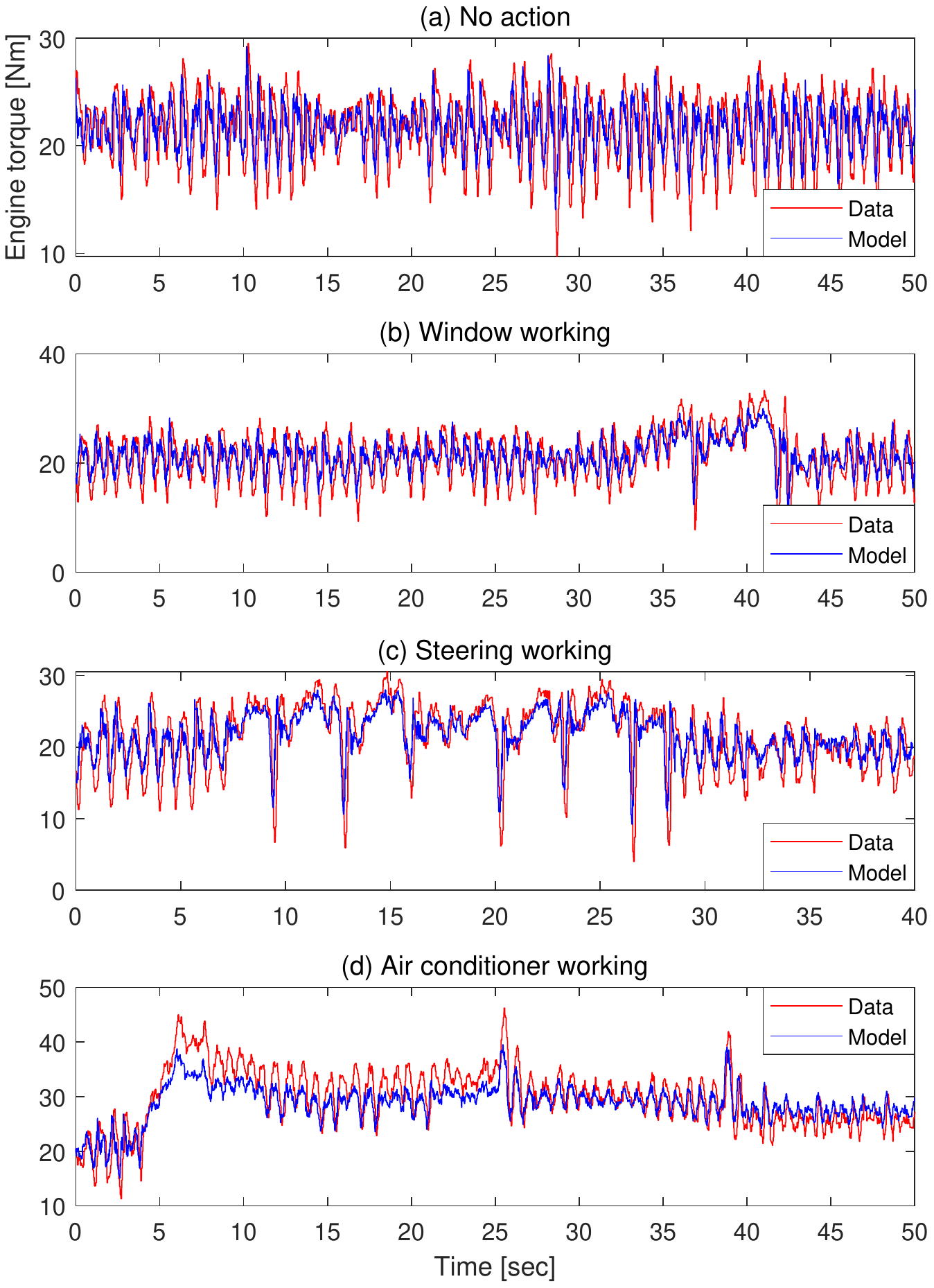}
\caption{Comparison of engine torque data and prediction from the model in (a) no action, (b) window working, (c) steering working, (d) air conditioner working cases.}
\label{fig2}
\end{center}
\end{figure}

Fig.~\ref{fig2} shows the entire prediction result of engine torque in each case from the constructed model with the estimated parameters. The training data set is the union of data in the interval 20$\sim$40 s for each case. The prediction from the model with the estimated parameters properly matches the overall data, though the prediction shows a little biased result from the real data in higher engine torque region in Fig.~\ref{fig2}(d). This prediction error occurs because the model with constant parameters cannot properly cover the higher engine torque region, but this slight prediction error due to model-plant mismatch can be handled at control system design stage using an offset-free control framework.

\subsection*{B. Air mass flow}
By combining (\ref{eq11}) and (\ref{eq12}), we can describe the air mass flow $w_{a,c}$ with intake manifold pressure $P_{im}$:
\begin{align}\label{eq17}
w_{a,c}(t)&=(\gamma_0+\gamma_1 N(t)+\gamma_2 N(t)^2)\\
&\frac{1}{\eta_c-1}\left( \eta_c-\left(\frac{P_e(t)}{P_{im}(t)}\right)^{1/\delta}\right)\frac{V_d\cdot P_{im}(t)\cdot N(t)}{2R\cdot T_{im}(t)}.\nonumber
\end{align}

Then, the parameters $\gamma_0,\gamma_1, \gamma_2$ and $\delta$ are estimated by solving the least mean squared error problem with $w_{a,c},P_{im}, T_{im}$ and $N$ data and given $\eta_c$ and $R$ values:
\begin{align}\label{eq18}
\underset{\small{\tilde{\gamma}_0, \tilde{\gamma}_1, \tilde{\gamma}_2,\delta}}{\mathrm{min}}\; \sum_{i=1}^{n_c} \sum_{j=1}^{n_{t}} ||w_{a,c}^{(i,j)}-\hat{w}_{a,c}^{(i,j)}||^2
\end{align}
where $w_{a,c}^{(i,j)}$ is $w_{a,c}$ data of case $i$ at the sampling time $j$, and $\hat{w}_{a,c}^{(i,j)}$ is the value calculated from (\ref{eq17}) with $P_{im}^{(i,j)}, T_{im}^{(i,j)}$ and $N^{(i,j)}$ data.

\begin{figure}[h]
\begin{center}
\includegraphics[width=9cm]{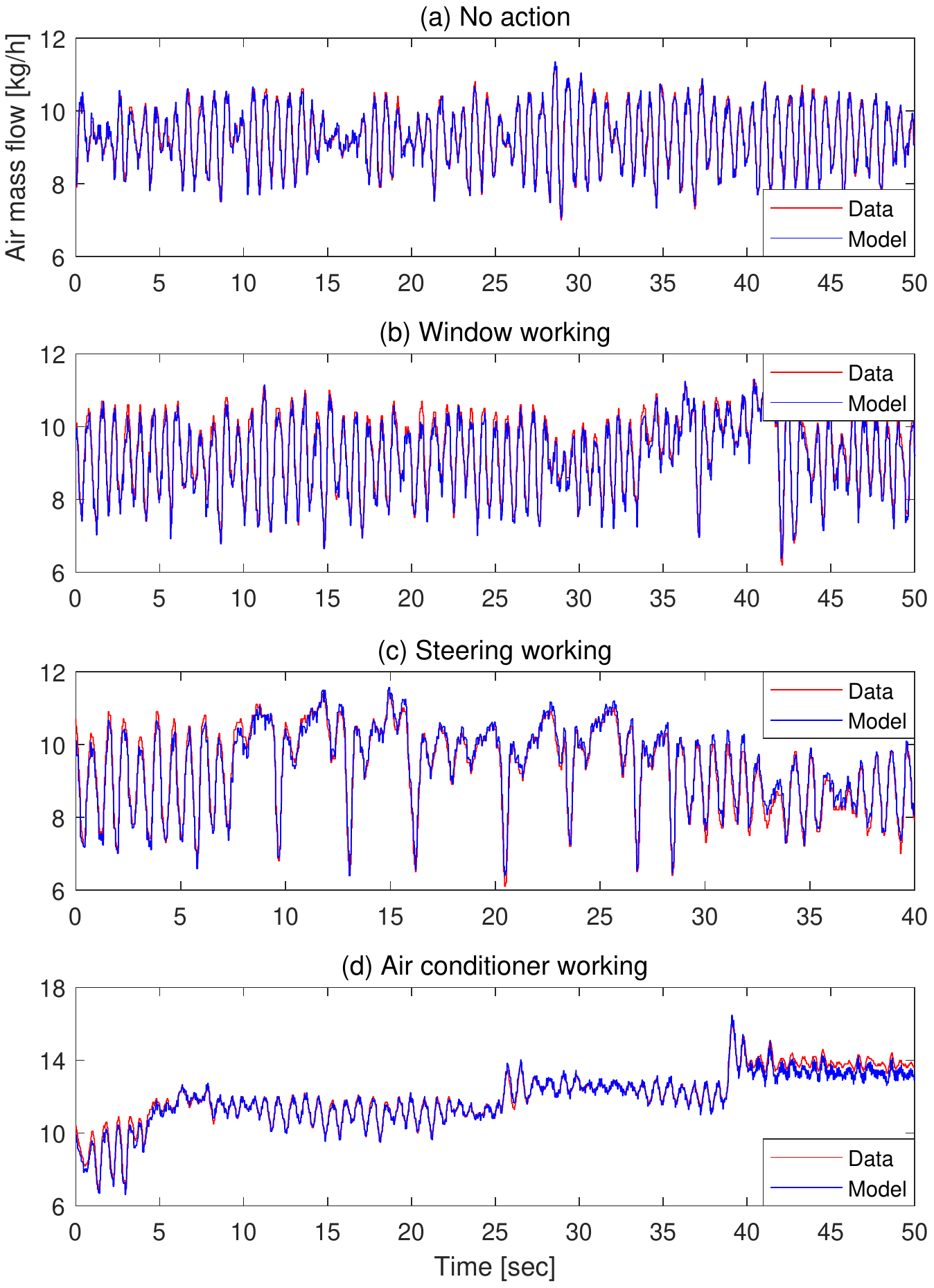}
\caption{Comparison of engine air mass flow data and prediction from the model in (a) no action, (b) window working, (c) steering working, (d) air conditioner working cases.}
\label{fig3}
\end{center}
\end{figure}

Fig.~\ref{fig3} shows the entire prediction result of air mass flow into the cylinder in each case from the constructed model with the estimated parameters. The training data set is the union of data in the interval 20$\sim$40 s for each case. Although the result in Fig.~\ref{fig3}(d) shows a little amount of prediction error in higher air-mass flow region, the values of predicted air mass flow with the model and estimated parameters properly match the overall air mass flow data.

In the case of throttle air mass flow $w_{a,th}$ model in (\ref{eq14}), the flow function $\mu$ can be approximated quite well with $\kappa=1.4$ for many working fluids such as intake air, exhaust gas at lower temperatures \cite{1}. Additionally, we used the given map of the engine to obtain throttle opening area $A_{th}$ for each throttle angle $\theta$.

\section{Control system design}\label{sec4}

\subsection*{A. Control-oriented model with system delay}\label{sec4a}

Substituting (\ref{eq15}) into (\ref{eq1}) yields:
\begin{align}\label{eq19}
\dot{N}(t)=c_1 \varepsilon_\zeta(t)\left(\beta_0+\beta_1 N(t)\right)\frac{w_{a,c}(t-\tau(t))}{N(t-\tau(t))}-c_2M_{loss}(t)
\end{align}
where $c_1:=60H_\ell/(\Theta_e\cdot\xi\cdot\eta)$ and $c_2:=60/\Theta_e$.

We set the engine speed $N$ as the state $\bar{x}\in\mathbb{R}^{n_x}$, the spark efficiency $\varepsilon_\zeta$ and cylinder air mass flow $w_{a,c}$ as the input $\bar{u}^\zeta\in\mathbb{R}^{n_{u}^\zeta}$ and $\bar{u}^w\in\mathbb{R}^{n_{u}^w}$, respectively, and torque loss $M_{loss}$ as the disturbance $\bar{d}\in\mathbb{R}^{n_d}$ in (\ref{eq19}):
\begin{align}\label{eq20}
\dot{\bar{x}}(t)=c_1 \bar{u}^\zeta(t)\left(\beta_0+\beta_1 \bar{x}(t)\right)\frac{\bar{u}^w(t-\tau(t))}{\bar{x}(t-\tau(t))}-c_2\bar{d}(t)
\end{align}
($n_x$, $n_{u^\zeta}$, $n_{u^w}$, and $n_d$ are 1 in this system, but we use those notations to provide a general formulation for low-complexity offset-free EMPC system in presence of system delay).

Then, we linearize (\ref{eq20}) based on values of each variable at the following  steady state:
\begin{align}
&\bar{x}_s=700 \:\mathrm{rpm},\; \bar{u}_{s}^\zeta=0.75,\; \bar{u}_{s}^w=9.239/3600\: \mathrm{kg/s},\nonumber\\
&\bar{d}_s=25\:\mathrm{Nm},\; \tau_s=60/700\:\mathrm{s}. \nonumber
\end{align}
The nominal spark efficiency value is usually set as 1 to increase the fuel efficiency, but in this study, the nominal spark efficiency value is set as a degraded value to utilize the spark efficiency as a manipulated variable. By this, the control system can reserve torque and compliment the slow actuation of air flow control as in \cite{2}.

We obtain a discrete-time model in (\ref{eq21}) by discretizing the linearized model with sampling instant 0.01 s:
\begin{align}\label{eq21}
&x(k+1)=Ax(k) +Bu^\zeta(k)+B_d d(k) \\
&\qquad\qquad+A_\tau x(k-\tau)+B_\tau u^w(k-\tau) \nonumber
\end{align}
where $x:=\bar{x}-\bar{x}_s$, $u^\zeta:=\bar{u}^\zeta-\bar{u}_s^\zeta$, $u^w:=\bar{u}^w-\bar{u}_s^w$, and $d:=\bar{d}-\bar{d}_s$. In this study, ISC is considered in the range of engine speed around the nominal idle speed, so a constant intake center to torque center delay is assumed, $\tau=9$.

To deal with the system delay, we introduced $e^d_{-i}$ to describe the effect of past state and input on current state:
\begin{align}\label{delay}
e^\tau_{-i}(k)=A_\tau x(k-i)+B_\tau u^w(k-i).
\end{align}
Then, by combining (\ref{eq21}) and (\ref{delay}), the prediction model in presence of system delay can be constructed as
\begin{align}
&\overbrace{\begin{bmatrix}
x(k+1) \\ e^{\tau}_{-1}(k+1) \\ e^{\tau}_{-2}(k+1) \\ \vdots \\ e^{\tau}_{-\tau}(k+1)\end{bmatrix}}^{x^e(k+1)}
=\overbrace{\begin{bmatrix}
A &\!\!\! 0 &\!\!\! \cdots &\!\!\! 0 & I \;\;\\
A_\tau &\!\!\! 0 &\!\!\! \cdots &\!\!\! 0 & 0 \;\;\\
0 &\!\!\! I &\!\!\! \cdots &\!\!\! 0 & 0 \;\;\\
\vdots &\!\!\! \vdots &\!\!\! \ddots &\!\!\! \vdots & \vdots \;\;\\
0 &\!\!\! 0 &\!\!\! \cdots &\!\!\! I & 0 \;\;\\
\end{bmatrix}}^{A^e}
\overbrace{\begin{bmatrix}
x(k) \\ e^{\tau}_{-1}(k) \\ e^{\tau}_{-2}(k) \\ \vdots \\ e^{\tau}_{-\tau}(k)\end{bmatrix}}^{x^e(k)}\nonumber\\
&\qquad\qquad\qquad\quad+ \overbrace{\begin{bmatrix} B &\!\!\! 0 \;\\ 0 &\!\!\! B_\tau \;\\ 0 &\!\!\! 0 \;\\ \vdots &\!\!\! \vdots \;\\ 0 &\!\!\! 0 \;\end{bmatrix}}^{B^e}
\overbrace{\begin{bmatrix} u^\zeta(k) \\ u^w(k) \end{bmatrix}}^{u(k)} \label{rev23}\\
&y(k)=\overbrace{\begin{bmatrix} I &\!\!\! 0 & \cdots &\!\!\! 0 \end{bmatrix}}^{C^e} x^e(k) \label{rev24}
\end{align}
where $x^e\in\mathbb{R}^{n_{x}^e}$ is the delay-augmented state, and $n_{x}^e$ denotes the dimension of $x^e$ ($n_{x}^e=(\tau+1)n_x$). In (\ref{rev23}) and (\ref{rev24}), the effect of the torque loss is not included. The torque loss is considered as a disturbance, and it is handled via offset-free MPC scheme in the next section.

\subsection*{B. Offset-free MPC system design}\label{sec4b}

We applied the standard offset-free MPC scheme with disturbance model and estimator which have been implemented in various processes \cite{45,46} to compensate for the effect of the torque loss $d$. In this scheme, we augment the linearized model with a disturbance model as
\begin{align}\label{eq23}
&\begin{cases} x^e(k+1)=A^e x^e(k) +B^e u(k) +B_{d}^e d(k) \\ d(k+1)=d(k) \\ y(k)=C^e x^e(k)+C_{d}^e d(k) \end{cases}
\end{align}
where $B_{d}^e\in\mathbb{R}^{n_{x}^e \times n_d}$ and $C_{d}^e\in\mathbb{R}^{n_y \times n_d}$ are matrices that represent the influence of the disturbance variable on the evolution of the delay-augmented state and the output, respectively.

To ensure the observability of the disturbance-augmented system in (\ref{eq23}), the original system (\ref{rev23}) and (\ref{rev24}) is observable and the following full column rank condition in (\ref{eq24}) should be satisfied \cite{12}.
\begin{align}\label{eq24}
{\mathrm{rank}} \begin{bmatrix} A^e-I & B_{d}^e \\ C^e & C_{d}^e \end{bmatrix} =n_{x}^e+n_d.
\end{align}

In this study, we set $B_{d}^e$ and $C_{d}^e$ as in (\ref{rev27}) and (\ref{rev28}) with $B_d$ in (\ref{eq21}) to make the disturbance variable $d$ have the same meaning as torque loss.
\begin{align}
&B_{d}^e:=[B_d^\top,0,\cdots,0]^\top \label{rev27}\\
&C_{d}^e:=\mathbf{0}_{n_y\times n_d}. \label{rev28}
\end{align}
These $B_{d}^e$ and $C_{d}^e$ in (\ref{rev27}) and (\ref{rev28}) also satisfy the condition in (\ref{eq24}). (Though $C_{d}^e$ is a zero matrix, we keep using that notation to provide a general formulation.)

Then, the delay-augmented state and disturbance estimator in (\ref{rev29}) is constructed based on the disturbance-augmented model in (\ref{eq23}).
\begin{align}\label{rev29}
\begin{bmatrix} \hat{x}(k+1)^e \\ \hat{d}(k+1) \end{bmatrix}&=
\begin{bmatrix} A^{e} & B_{d}^e \\ 0 & I \end{bmatrix} \begin{bmatrix} \hat{x}^e(k) \\ \hat{d}(k) \end{bmatrix}+ \begin{bmatrix} B^{e} \\ 0 \end{bmatrix} u(k) \nonumber\\
& + \begin{bmatrix} L_{x}^e \\ L_d \end{bmatrix} (-y_{m}(k)+C^{e}\hat{x}^e(k)+C_{d}^e\hat{d}(k))
\end{align}
where $L_{x}^e\in\mathbb{R}^{n_{x}^e \times n_y}$ and $L_d\in\mathbb{R}^{n_d \times n_y}$ are the estimator gains for the delay-augmented state and the disturbance, respectively, that make the estimator stable, and $y_m$ is the output measurement.

Given the estimated delay-augmented state and disturbance from the estimator in (\ref{rev29}), a finite-horizon optimal control problem in (\ref{rev30}) is solved to obtain the optimal spark efficiency and cylinder air flow that drives the engine speed to the desired idle speed under the influence of disturbance (i.e., torque loss) \cite{13,14,21}.
\begin{subequations}\label{rev30}
\begin{align}
\underset {u_{i},\epsilon}{\mathrm{min}}\quad &\sum_{i=0}^{N-1}||y_{i+1}-r_y||^2_{Q_y}+||u_{i}^\zeta-r_{u}^\zeta||^2_{Q_{u}^\zeta}+||\delta u_{i}||^2_{Q_{\delta u}}\nonumber\\
&\qquad+||y_N-r_y||^2_{Q^N_y} + Q_\epsilon \epsilon^2\label{rev30a}\\
\mathrm{s.t.}\quad & x_{0}^e=\hat{x}^e(k),\; d=\hat{d}(k),\; u_{-1}=u(k-1) \label{rev30b}\\
&x_{i+1}^e=A^e x_{i}^e +B^e u_{i} +B_{d}^e d\label{rev30c}\\
&y_i=C^e x_{i}^e+C_{d}^e d \label{rev30d}\\
&u_{\mathrm{min}}\leq u_{i}\leq u_{\mathrm{max}}\label{rev30e}\\
&\delta u_{\mathrm{min}}\leq \delta u_{i}\leq \delta u_{\mathrm{max}}\label{rev30f}\\
&y_\mathrm{min}-\epsilon\mathbf{1}\leq y_i\leq y_\mathrm{max}+\epsilon\mathbf{1} \label{rev30g}\\
&\epsilon\ge 0,\quad i=0,\cdots,N-1 \label{rev30h}
\end{align}
\end{subequations}
where $r_y:=\bar{r}_y-\bar{x}_s$ and $r_{u}^\zeta:=\bar{r}_{u}^\zeta-\bar{u}_s^\zeta$. $\bar{r}_y$ and $\bar{r}_{u}^\zeta$ denote the set-point values of engine speed and spark efficiency for torque reserve, respectively, $\delta u_i$ denotes the input variation, and $\epsilon$ denotes the slack variable to apply the soft constraint to the output constraint in (\ref{rev30g}). The objective function in (\ref{rev30a}) is set to regulate the engine speed for ISC and the spark efficiency for torque reserve while minimizing the input variation and the slack variable in the soft constraint.

\subsection*{C. Low-complexity EMPC formulation}\label{sec4c}

Since the capacity assigned for the ECU is limited, a direct implementation of optimization algorithm into the ECU is not available. Therefore, we apply the methodology of explicit model predictive control \cite{15,16}, which derives the explicit solution map by solving multiparametric program off-line and allows to obtain an optimal solution on-line without solving optimization problem, to the engine speed control system. Since EMPC obtains an optimal solution within a considerably short time by moving the computational effort for on-line optimization to off-line, it is also ideal to make the engine speed control system computationally feasible where the sampling interval is quite short.  

Moreover, we develop a low-complexity mp-QP formulation with constraint horizon in presence of system delay in input and state variables based on the basic parametric program schemes \cite{17,18} to decrease the complexity of the resultant explicit solution map. By this, the computational burden for on-line evaluation and the memory consumption in ECU can be further reduced. The detailed flow of the low-complexity mp-QP formulation is described below.

\begin{figure*}[h]
\begin{center}
\includegraphics[height=3.5cm]{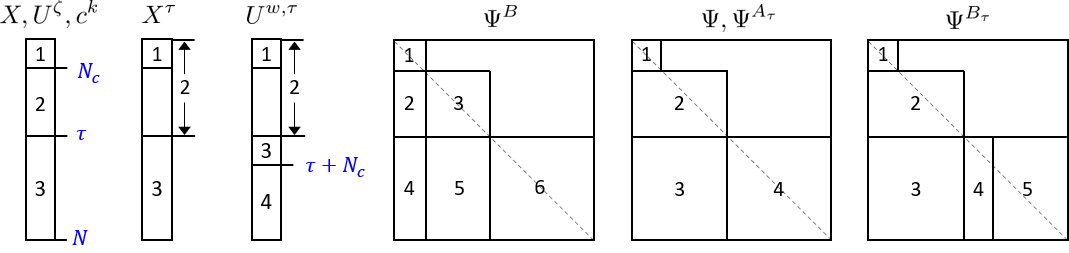}
\caption{Partitions of system vectors and matrices.}
\label{fig4}
\end{center}
\end{figure*}

The future state prediction from the dynamics in (\ref{eq21}) can be formulated as in (\ref{eq26}).
\begin{align}\label{eq26}
&\!\!\!\!\!\!\! \overbrace{\begin{bmatrix}x_{1} \\ x_{2} \\ \vdots \\ x_{N} \end{bmatrix}}^{X}
=\overbrace{\begin{bmatrix} A \\ A^2 \\ \vdots \\ A^{N} \end{bmatrix}}^{\Phi} x_0 
+\overbrace{\begin{bmatrix} B_d &\!\!\! 0 &\!\!\! \cdots &\!\!\! 0 \\
AB_d &\!\!\! B_d &\!\!\! \cdots &\!\!\! 0 \\
\vdots & \vdots & \ddots & \vdots \\
A^{\textit{N}-1}B_d &\!\!\! A^{\textit{N}-2}B_d &\!\!\! \cdots &\!\!\! B_d \end{bmatrix}}^{\Psi^{B_d}}
\begin{bmatrix}d \\ d \\ \vdots \\ d \end{bmatrix} \nonumber\\
&\quad +\overbrace{\begin{bmatrix} B &\!\!\! 0 &\!\!\! \cdots &\!\!\! 0 \\ 
AB &\!\!\! B &\!\!\! \cdots &\!\!\! 0 \\ 
\vdots & \vdots & \ddots & \vdots \\ 
A^{\textit{N}-1}B &\!\!\! A^{\textit{N}-2}B &\!\!\! \cdots &\!\!\! B \end{bmatrix}}^{\Psi^B}
\overbrace{\begin{bmatrix}u_{0}^\zeta \\ u_{1}^\zeta \\ \vdots \\ u_{N-1}^\zeta \end{bmatrix}}^{U^\zeta} \nonumber\\
&\quad+\overbrace{\begin{bmatrix} A_\tau &\!\!\! 0 &\!\!\! \cdots &\!\!\! 0 \\ 
AA_\tau &\!\!\! A_\tau &\!\!\! \cdots &\!\!\! 0 \\ 
\vdots & \vdots & \ddots & \vdots \\ 
A^{\textit{N}-1}A_\tau &\!\!\! A^{\textit{N}-2}A_\tau &\!\!\! \cdots &\!\!\! A_\tau \end{bmatrix}}^{\Psi^{A_\tau}}
\overbrace{\begin{bmatrix}x_{-\tau} \\ x_{-\tau+1} \\ \vdots \\ x_{-\tau+N-1} \end{bmatrix}}^{X^\tau} \nonumber\\
&\quad+\overbrace{\begin{bmatrix} B_\tau &\!\!\! 0 &\!\!\! \cdots &\!\!\! 0 \\ 
AB_\tau &\!\!\! B_\tau &\!\!\! \cdots &\!\!\! 0 \\ 
\vdots & \vdots & \ddots & \vdots \\ 
A^{\textit{N}-1}B_\tau &\!\!\! A^{\textit{N}-2}B_\tau &\!\!\! \cdots &\!\!\! B_\tau \end{bmatrix}}^{\Psi^{B_\tau}}
\overbrace{\begin{bmatrix}u_{-\tau}^w \\ u_{-\tau+1}^w \\ \vdots \\ u_{-\tau+N-1}^w \end{bmatrix}}^{U^{w,\tau}}.
\end{align}
We can describe (\ref{eq26}) simply as
\begin{align}\label{eq27}
&X=\ell^{0}+\Psi^B U^\zeta +\Psi^{A_\tau} X^\tau+\Psi^{B_\tau} U^{w,\tau} \\
&\ell^{0}:=\Phi x_0 +\Psi^{B_d} (\textbf{1}_N\otimes d). \nonumber
\end{align}
The predicted future output can be described as
\begin{align}\label{eq28}
&Y=\mathbf{C}_y X+\mathbf{C}_d d\\
&\mathbf{C}_y := I_{N} \otimes C^e,\; \mathbf{C}_d := \mathbf{1}_{N} \otimes C_d^e \nonumber
\end{align}
where $\mathbf{1}_N$ represents a vector of ones with length $N$.

Now, we divide the vectors of variables and system matrices within the prediction horizon $N$ into several parts considering constraint horizon $N_c$ and system delay $\tau$ as shown in Fig.~\ref{fig4}. 
Then, the divided future states can be reformulated as (\ref{eq29a})--(\ref{eq29c}).
\begin{align}
&\!\!\!X_1=\ell^0_1 +\Psi_1^{A_\tau} X_1^\tau +\Psi_1^{B_\tau} U_1^{w,\tau} + \Psi_1^B U_1^\zeta \label{eq29a}\\
&\!\!\!X_2=\ell^0_2 +\Psi_2^{A_\tau} X_2^\tau +\Psi_2^{B_\tau} U_2^{w,\tau} + \Psi_2^B U_1^\zeta + \Psi_3^B U_2^\zeta \label{eq29b}\\
&\!\!\!X_3=\ell^0_3 +\Psi_3^{A_\tau} X_2^\tau +\Psi_3^{B_\tau} U_2^{w,\tau} + \Psi_4^B U_1^\zeta + \Psi_5^B U_2^\zeta + \Psi_6^B U_3^\zeta \nonumber\\
&\qquad\quad + \Psi_4^{A_\tau} X_3^\tau + \Psi_4^{B_\tau} U_3^{w,\tau} +\Psi_5^{B_\tau} U_4^{w,\tau}\label{eq29c}
\end{align}
where subscript numbers denote the partition numbers of each variable and matrix illustrated in Fig.~\ref{fig4}. 

$X_1$ and $X_2$ can be easily rearranged by separating the terms that are independent of future input values as in (\ref{rev37}) and (\ref{rev38}).
\begin{align}
&X_1=\ell_{X_1}+\Psi_1^B U_1^\zeta \label{rev37}\\
&X_2=\ell_{X_2}+\Psi_2^B U_1^\zeta +\Psi_3^B U_2^\zeta \label{rev38}
\end{align}
where $\ell_{X_1}$ and $\ell_{X_2}$ denote the terms independent of future input values of $X_1$ and $X_2$, respectively:
\begin{align}
&\ell_{X_1}:=\ell^0_1 +\Psi_1^{A_\tau} X_1^\tau +\Psi_1^{B_\tau} U_1^{w,\tau} \label{new39}\\
&\ell_{X_2}:=\ell^0_2 +\Psi_2^{A_\tau} X_2^\tau +\Psi_2^{B_\tau} U_2^{w,\tau}. \label{new40}
\end{align}

In the case of $X_3$, since $X_3^\tau$ term contains future states unlike $X_1^\tau$ and $X_2^\tau$ that only consist of past state values, we have to consider the future-input dependent part in $X_3^\tau$. $X_3^\tau$ can be reformulated with the future state partition $X_1$ and $X_2$ as in (\ref{rev39}).
\begin{align}\label{rev39}
&\qquad\qquad\qquad X_3^\tau = C^1_x x_0 + C^2_x X_1+ C^3_x X_2\\
&C_x^1:= \begin{bmatrix} I_{n_x} \\ \mathbf{0}_{(N-\tau-1)n_x,n_x} \end{bmatrix} ,\;
C^2_x:= \begin{bmatrix} \mathbf{0}_{n_x,N_{c} n_x} \\ I_{N_{c} n_x} \\ \mathbf{0}_{(N-\tau-N_{c}-1)n_x,N_{c} n_x} \end{bmatrix} \nonumber\\
&C^3_x:= \begin{bmatrix} \mathbf{0}_{(N_{c}+1)n_x,(N-\tau-N_{c}-1) n_x}\!\!\! & \!\!\!\mathbf{0}_{(N_{c}+1)n_x,(2\tau-N+1) n_x} \\ I_{(N-\tau-N_{c}-1) n_x}\!\!\! & \!\!\!\mathbf{0}_{(N-\tau-N_{c}-1) n_x,(2\tau-N+1) n_x} \\ \end{bmatrix}. \nonumber
\end{align}
Then, substituting (\ref{rev37}) and (\ref{rev38}) into (\ref{rev39}) and rearranging yields
\begin{align}\label{rev40}
&X_3^\tau=C^1_x x_0 +C^2_x \ell_{X_1}+ C^3_x \ell_{X_2} \nonumber\\
&\qquad +(C^2_x \Psi_1^B + C^3_x \Psi_2^B)U_1^\zeta + C^3_x \Psi_3^B U_2^\zeta.
\end{align}
Now, by substituting (\ref{rev40}) into (\ref{eq29c}) and rearranging, we can reformulate $X_3$ as 
\begin{align}
&X_3=\ell_{X_3} +\Psi_{U_1^\zeta}^{X_3}U_1^\zeta +\Psi_{U_2^\zeta}^{X_3}U_2^\zeta +\Psi_6^B U_3^\zeta \label{rev41}\\
&\qquad\qquad+\Psi_4^{B_\tau}U_3^{w,\tau} +\Psi_5^{B_\tau} U_4^{w,\tau} \nonumber\\
&\Psi_{U_1^\zeta}^{X_3}:=\Psi_4^{A_\tau}C^2_x \Psi_1^B +\Psi_4^{A_\tau}C^3_x \Psi_2^B +\Psi_4^B \nonumber\\
&\Psi_{U_2^\zeta}^{X_3}:=\Psi_4^{A_\tau}C^3_x \Psi_3^B+\Psi_5^B.\nonumber
\end{align}
where $\ell_{X_3}$ is the future-input dependent part of $X_3$:
\begin{align}\label{new44}
&\ell_{X_3}:=\ell^0_3 +\Psi_3^{A_\tau} X_2^\tau +\Psi_3^{B_\tau} U_2^{w,\tau}\\
&\qquad\qquad+ \Psi_4^{A_\tau}(C^1_x x_0 +C^2_x \ell_{X_1}+ C^3_x \ell_{X_2}) \nonumber.
\end{align}

By substituting (\ref{rev37}), (\ref{rev38}), and (\ref{rev41}) into (\ref{eq28}), we can describe the predicted future output with the constant term dependent on the past variables $X^\tau_1,X^\tau_2,U^{w,\tau}_1$ and $U^{w,\tau}_2$ and the term dependent on future inputs $U_1^\zeta,U_2^\zeta,U_3^\zeta,U_3^{w,\tau}$ and $U_4^{w,\tau}$. 

Now, we separate the future inputs into variables within and outside the constraint horizon $N_c$:
\begin{align}
&\mathbf{U}_{c}:=[U_1^{\zeta\top},U_3^{w,\tau\top}]^\top \label{rev42}\\
&\mathbf{U}_{\bar{c}}:=[U_2^{\zeta\top},U_3^{\zeta\top},U_4^{w,\tau\top}]^\top \label{rev43}
\end{align}
where $\mathbf{U}_{c}$ denotes the future inputs within $N_c$, and $\mathbf{U}_{\bar{c}}$ denotes the future inputs outside $N_c$. Then, we describe the objective function in (\ref{rev30a}) and constraints in (\ref{rev30e})--(\ref{rev30h}) with $\mathbf{U}_{c}$ to reconstruct the optimal control problem (\ref{rev30}) in a compact form with $\mathbf{U}_{c}$.

The objective function in (\ref{rev30a}) can be reformulated as 
\begin{align}\label{rev44}
&\qquad\quad  J=||Y-\mathbf{r}_y||^2_{Q_Y} +||U^\zeta-\mathbf{r}_u^\zeta||^2_{Q_U^\zeta} \nonumber\\
&\qquad\qquad+||\Delta U^\zeta||^2_{Q_{\Delta U}^\zeta} +||\Delta U^{w}||^2_{Q^w_{\Delta U}} +Q_\epsilon \epsilon^2 \\
&Q_Y:=diag\{ Q_y,\cdots,Q_y,Q_y^N \},\;Q_U^\zeta:=diag\{ Q_u,\cdots,Q_u \}\nonumber\\
&Q_{\Delta U}^\zeta:=diag\{ Q_{\delta u}^\zeta,\cdots,Q_{\delta u}^\zeta \},\; Q_{\Delta U}^{w}:=diag\{ Q_{\delta u}^{w},\cdots, Q_{\delta u}^{w} \} \nonumber\\
&\mathbf{r}_y:=\mathbf{1}_N \otimes r_y,\;\mathbf{r}_u^\zeta:=\mathbf{1}_N \otimes r_u^\zeta \nonumber
\end{align}
where $\Delta$ represents the variation of variables, and $U^{w}$ denotes the future cylinder air flow $\begin{bmatrix} U_{3}^{w,\tau\top},U_{4}^{w\top} \end{bmatrix}^\top$. We can also reformulate $Y,U^\zeta,U^{w},\Delta U^\zeta$ and $\Delta U^{w}$ with $\mathbf{U}_{c}$ and $\mathbf{U}_{\bar{c}}$ as
\begin{align}
&Y = \ell_Y+S^Y_{c}\mathbf{U}_{c}+ S^Y_{\bar{c}}\mathbf{U}_{\bar{c}} \label{eq32}\\
&U^\zeta = S^\zeta_{c} \mathbf{U}_{c}+ S^\zeta_{\bar{c}} \mathbf{U}_{\bar{c}} \label{eq33}\\
&U^{w} = S^{w}_{c} \mathbf{U}_{c}+ S^{w}_{\bar{c}} \mathbf{U}_{\bar{c}}\\
&\Delta U^\zeta = S^{\Delta \zeta}_{c} \mathbf{U}_{c}+ S^{\Delta \zeta}_{\bar{c}} \mathbf{U}_{\bar{c}}-\mathbf{u}_{-1}^\zeta \label{eq34}\\
&\Delta U^{w} = S^{\Delta w}_{c} \mathbf{U}_{c}+ S^{\Delta w}_{\bar{c}} \mathbf{U}_{\bar{c}}-\mathbf{u}_{-1}^w\label{eq35}
\end{align}
where
\begin{align}
&S^Y_{c}:=\mathbf{C}_y \begin{bmatrix} \Psi_1^B & 0\\ \Psi_2^B & 0\\ \Psi^{X_3}_{U_1} & \Psi^{B_\tau}_4\end{bmatrix} \nonumber\\
&S^Y_{\bar{c}}:= \mathbf{C}_y \begin{bmatrix} 0 & 0 & 0 \\ \Psi_3^B & 0 & 0 \\ \Psi^{X_3}_{U_2} & \Psi_6^B & \Psi_5^{B_\tau} \end{bmatrix},\;S^\zeta_{c}:=\begin{bmatrix} I & 0\\ 0 & 0\end{bmatrix} \nonumber \\
&S^\zeta_{\bar{c}}:=\begin{bmatrix} 0 & 0 & 0 \\ I & 0 & 0 \\ 0 & I & 0 \end{bmatrix},\; S^{w}_{c}:=\begin{bmatrix} 0 & I \\ 0 & 0\end{bmatrix},\; S^{w}_{\bar{c}}:=\begin{bmatrix} 0 & 0 & 0 \\ 0 & 0 & I \end{bmatrix}\nonumber\\
&S^{\Delta \zeta}_{c}:=C^{N,n_u}_\Delta S^\zeta_{c},\; S^{\Delta \zeta}_{\bar{c}}:=C^{N,n_u}_\Delta S^\zeta_{\bar{c}}  \nonumber\\
&S^{\Delta w}_{c}:=C^{N-\tau,n_u^\tau}_\Delta S^{w}_{c},\; S^{\Delta w}_{\bar{c}}:=C^{N-\tau,n_u^\tau}_\Delta S^{w}_{\bar{c}}\nonumber\\
&C_\Delta^{n,m} :=\overbrace{\begin{bmatrix}
I_m &\!\!\! 0 &\!\!\! 0 &\!\!\! \cdots &\!\!\! 0 &\!\!\! 0\;\;\\
-I_m &\!\!\! I_m &\!\!\! 0 &\!\!\! \cdots &\!\!\! 0 &\!\!\! 0\;\;\\
0 &\!\!\!\!\!\! -I_m &\!\!\! I_m &\!\!\! \cdots &\!\!\! 0 &\!\!\! 0\;\;\\
\vdots &\!\!\! \vdots &\!\!\! \vdots & \ddots &\!\!\! \vdots &\!\!\! \vdots\;\; \\
0 &\!\!\! 0 &\!\!\! 0 &\!\!\! \cdots &\!\!\!\!\! -I_m &\!\!\! I_m\;\;
\end{bmatrix}}^{n} \nonumber\\
&\mathbf{u}_{-1}^\zeta:= \begin{bmatrix} {u_{-1}^{\zeta\top}},0,\cdots,0 \end{bmatrix}^\top,\;\mathbf{u}_{-1}^w:= \begin{bmatrix} {u^{w\top}_{-1}},0,\cdots,0 \end{bmatrix}^\top.\nonumber
\end{align}
$\ell_Y$ is the future-input independent term of $Y$ in (\ref{eq32}).
\begin{align}\label{new53}
\ell_Y:=\mathbf{C}_y \begin{bmatrix} \ell_{X_1} \\ \ell_{X_2} \\ \ell_{X_3} \end{bmatrix} +\mathbf{C}_d d.
\end{align}
Then, by substituting (\ref{eq32})--(\ref{eq35}) into (\ref{rev44}) and rearranging, $J$ can be reformulated in a quadratic form of $\mathbf{U}_{\bar{c}}$ as in (\ref{rev50}).
\begin{align}\label{rev50}
&J=\mathbf{U}_{\bar{c}}^\top H_{\bar{c}} \mathbf{U}_{\bar{c}} +2\mathbf{U}_{\bar{c}}^\top f_{\bar{c}} +\textbf{U}_{c}^\top H_{c} \textbf{U}_{c}+2\textbf{U}_{c}^\top f_{c} +Q_\epsilon \epsilon^2 +c_J \\
&H_{\bar{c}}:=S^{Y\top}_{\bar{c}} Q_Y S^Y_{\bar{c}} +S^{\zeta\top}_{\bar{c}} Q_U^\zeta S^\zeta_{\bar{c}} \nonumber\\
&\qquad\quad +S^{\Delta\zeta\top}_{\bar{c}}Q_{\Delta U}^\zeta S^{\Delta\zeta}_{\bar{c}} +S^{\Delta w\top}_{\bar{c}} Q_{\Delta U}^w S^{\Delta w}_{\bar{c}}\nonumber\\
&f_{\bar{c}}:=S^{Y\top}_{\bar{c}} Q_Y (\ell_Y+S^Y_c \mathbf{U}_{c}-\mathbf{r}_y) +S^{\zeta\top}_{\bar{c}} Q_U^\zeta (S^\zeta_c \mathbf{U}_c -\mathbf{r}_u^\zeta)\nonumber\\
&\;+S^{\Delta \zeta\top}_{\bar{c}} Q_{\Delta U}^\zeta (S^{\Delta\zeta}_c \mathbf{U}_c -\mathbf{u}_{-1}^\zeta) +S^{\Delta w\top}_{\bar{c}} Q_{\Delta U}^w (S^{\Delta w}_{c}\mathbf{U}_c-\mathbf{u}_{-1}^w)\nonumber
\end{align}
where $c_J$ denotes the constant term, and $H_{c}$ and $f_c$ are
\begin{align}
&H_{c}:=S^{Y\top}_c Q_Y S^Y_c +S^{\zeta\top}_c Q_U^\zeta S^\zeta_c  \label{new55}\\
&\qquad\quad +S^{\Delta\zeta\top}_{c}Q_{\Delta U}^\zeta S^{\Delta\zeta}_{c} +S^{\Delta w\top}_{c} Q_{\Delta U}^w S^{\Delta w}_{c}\nonumber\\
&f_{c}:=S^{Y\top}_c Q_Y(\ell_Y-\mathbf{r}_{y})-S^{\zeta\top}_c Q_U^\zeta \mathbf{r}_{u}^\zeta \label{new56}\\
&\quad\qquad - S^{\Delta\zeta\top}_{c} Q_{\Delta U}^\zeta \mathbf{u}_{-1}^\zeta -S^{\Delta w\top}_{c} Q_{\Delta U}^w \mathbf{u}_{-1}^w.\nonumber
\end{align}
Since $\mathbf{U}_{\bar{c}}$ does not affect the future variables within the constraint horizon $\mathbf{U}_{c}$, we can analytically derive the unconstrained optimal solution $\mathbf{U}_{\bar{c}}^*$ that minimizes the value of $J$:
\begin{align}\label{rev51}
\mathbf{U}_{\bar{c}}^*=-H_{\bar{c}}^{-1} f_{\bar{c}}.
\end{align}
Since $f_{\bar{c}}$ is dependent of $\mathbf{U}_c$, $\mathbf{U}_{\bar{c}}^*$ is a predetermined function of $\mathbf{U}_c$. Then, by substituting (\ref{rev51}) into the objective function in (\ref{rev50}) and rearranging, the objective function can be reformulated more compactly as a function of $\mathbf{U}_{c,\epsilon}:=\begin{bmatrix} \mathbf{U}_c^\top,\epsilon \end{bmatrix}^\top$:
\begin{align}\label{rev52}
&\overline{J}=\textbf{U}_{c,\epsilon}^\top \bar{H} \textbf{U}_{c,\epsilon} +2\textbf{U}_{c,\epsilon}^\top \bar{f}+\bar{c}_J\\
&\bar{H}:=\begin{bmatrix} \bar{H}_c & 0 \\ 0 & Q_\epsilon \end{bmatrix},\; \bar{f}:=\begin{bmatrix} \bar{f}_c & 0 \end{bmatrix} \nonumber
\end{align}
where $\bar{c}_J$ denotes the constant term, and $\bar{H}_c$ and $\bar{f}_c$ are
\begin{align}
&\bar{H}_c:=H_c-S^{f_{\bar{c}}\top}_{c}H_{\bar{c}}^{-1}S^{f_{\bar{c}}}_{c} \label{new59}\\
&\bar{f}_c:=f_c-S^{f_{\bar{c}}\top}_{c}H_{\bar{c}}^{-1}\ell_{f_{\bar{c}}} \label{new60}\\
&S^{f_{\bar{c}}}_{c}:=S^{Y\top}_{\bar{c}} Q_Y S^Y_c +S^{\zeta\top}_{\bar{c}} Q_U^\zeta S^\zeta_c \label{renew61}\\
&\quad\qquad +S^{\Delta\zeta\top}_{\bar{c}}Q_{\Delta U}^\zeta S^{\Delta\zeta}_c +S^{\Delta w\top}_{\bar{c}}Q_{\Delta U}^w S^{\Delta w}_c \nonumber\\
&\ell_{f_{\bar{c}}}:=S^{Y\top}_{\bar{c}} Q_Y (\ell_Y-\mathbf{r}_y) -S^{\zeta\top}_{\bar{c}} Q_U^\zeta \mathbf{r}_{u}^\zeta \label{renew62}\\
&\quad\qquad - S^{\Delta\zeta\top}_{\bar{c}} Q_{\Delta U}^\zeta \mathbf{u}_{-1}^\zeta -S^{\Delta w\top}_{\bar{c}} Q_{\Delta U}^w\mathbf{u}_{-1}^w  \nonumber
\end{align}
By applying (\ref{rev52}) to (\ref{rev30}), we can effectively decrease the complexity of the problem by reducing the number of variables.

In the case of constraints in (\ref{rev30e})--(\ref{rev30g}), they can also be reformulated with $\mathbf{U}_c$ by applying the constraint horizon. The constraints on input in (\ref{rev30e}) and input variation in (\ref{rev30f}) are easily reformulated as
\begin{align}
&\begin{bmatrix}  I_{N_{c}}\otimes F_{\zeta} &\!\!\! 0 \\ 0 &\!\!\! I_{N_{c}}\otimes F_{w} \end{bmatrix} \mathbf{U}_c \leq \begin{bmatrix} \mathbf{1}_{N_{c}}\otimes g_{\zeta} \\ \mathbf{1}_{N_{c}}\otimes g_{w} \end{bmatrix} \label{rev53}\\
&\begin{bmatrix}  I_{N_{c}}\otimes F_{\delta \zeta} &\!\!\! 0 \\ 0 &\!\!\! I_{N_{c}}\otimes F_{\delta w} \end{bmatrix} \begin{bmatrix}  C^{N_{c},n_{u}^\zeta}_\Delta &\!\!\! 0 \\ 0 &\!\!\! C^{N_{c},n_{u}^w}_\Delta \end{bmatrix}\mathbf{U}_c \label{rev54}\\
&\quad\leq \begin{bmatrix} \mathbf{1}_{N_{c}}\otimes g_{\delta\zeta} \\ \mathbf{1}_{N_{c}}\otimes g_{\delta w} \end{bmatrix} +\begin{bmatrix}  I_{N_{c}}\otimes F_{\delta \zeta} &\!\!\! 0 \\ 0 &\!\!\! I_{N_{c}}\otimes F_{\delta w} \end{bmatrix} \begin{bmatrix} \mathbf{u}_{-1}^{\zeta,N_{c}} \\ \mathbf{u}_{-1}^{w,N_{c}} \end{bmatrix} \nonumber
\end{align}
where
\begin{align}
&F_{\zeta},F_{\delta\zeta}:=\begin{bmatrix} -I_{n_{u}^\zeta} \\ I_{n_{u}^\zeta} \end{bmatrix},\; F_{w},F_{\delta w}:=\begin{bmatrix} -I_{n_{u}^w} \\ I_{n_{u}^w} \end{bmatrix}\nonumber\\
&g_{\zeta}:=\begin{bmatrix} -{u}^\zeta_{min} \\ {u}^\zeta_{max} \end{bmatrix},\; g_{w}:=\begin{bmatrix} -{u}^w_{min} \\ {u}^w_{max} \end{bmatrix}\nonumber\\
&g_{\delta\zeta}:=\begin{bmatrix} -{\delta u}^\zeta_{min} \\ {\delta u}^\zeta_{max} \end{bmatrix},\; g_{\delta w}:=\begin{bmatrix} -{\delta u}^w_{min} \\ {\delta u}^w_{max} \end{bmatrix}\nonumber\\
&\mathbf{u}_{-1}^{\zeta,N_{c}}= \begin{bmatrix} u_{-1}^\zeta\\ \mathbf{0}_{N_{c}-1}\end{bmatrix},\; \mathbf{u}_{-1}^{w,N_{c}}= \begin{bmatrix} u_{-1}^w\\ \mathbf{0}_{N_{c}-1}\end{bmatrix} \nonumber
\end{align}
The constraint on future output in (\ref{rev30g}) can be reformulated by substituting the first $N_c$ components of $Y$ (\ref{eq32}) into (\ref{rev30g}) and rearranging with $\mathbf{U}_c$:
\begin{align}
(I_{N_{c}}\otimes F_y) &\mathbf{C}_y^{N_{c}} \begin{bmatrix} \Psi_1^B &\!\!\! 0 \end{bmatrix} \mathbf{U}_c +(\mathbf{1}_{N_{c}}\otimes F_y^\epsilon) \epsilon \label{rev55}\\
& \leq\mathbf{1}_{N_{c}}\otimes g_y -(I_{N_{c}}\otimes F_y)\ell_{Y}^{N_c} \nonumber\
\end{align}
where
\begin{align}
&F_{y}:=\begin{bmatrix} -I_{n_{y}} \\ I_{n_{y}} \end{bmatrix},\; g_{y}:=\begin{bmatrix} -{y}_{min} \\ {y}_{max} \end{bmatrix},\; F_y^\epsilon:=\begin{bmatrix} -\mathbf{1}_{n_y} \\ \mathbf{1}_{n_y} \end{bmatrix} \nonumber\\
&\ell_{Y}^{N_c}:=\mathbf{C}_y^{N_{c}} \ell_{X_1} +\mathbf{C}_d^{N_{c}} d \nonumber\\
&\mathbf{C}_y^{N_c} := I_{N_c} \otimes C^e,\; \mathbf{C}_d^{N_c} := \mathbf{1}_{N_c} \otimes C_d^e. \nonumber
\end{align}

Now, we select the parameters for parametric programming. In the case of the objective function in (\ref{rev52}), $\bar{H}_c$ and $Q_\epsilon$ are constant matrices. On the other hand, $\bar{f}_c$ depends on current state $x_k$, disturbance $d_k$, output reference $r_y$, input reference $r_u^\zeta$, past states $x_{k-1},\cdots,x_{k-\tau}$, and past inputs $u_{k-1},\cdots,u_{k-\tau}$. Since considering all the variables described above as parameters is inefficient, $\bar{f}_c$ itself is set as a parameter. In the case of constraints in (\ref{rev53})--(\ref{rev55}), $ u_{-1}^\zeta,u_{-1}^w,\ell_{Y}^{N_c}$ are selected as parameters to specify the inequalities. Consequently, by considering the parameters in (\ref{rev56}), we can efficiently specify the optimal control problem of the developed low-complexity offset-free MPC with constraint horizon in presence of system delay.
\begin{align}
p=[\bar{f}^\top_c,u_{-1}^{\zeta\top},u_{-1}^{w\top},\ell_{Y}^{N_c\top}]^\top\label{rev56}
\end{align}

The parameters $\bar{f}_c$ and $\ell_{Y}^{N_c}$ can be derived from $x_0^e:=[x_0^\top,e_{-1}^{\tau\top},\cdots,e_{-\tau}^{\tau\top}]$ and $d$ by the following procedure. First, $\ell_{X_1}$, $\ell_{X_2}$, and $\ell_{X_3}$ in (\ref{new39}), (\ref{new40}), and (\ref{new44}) can be obtained from $x_0^e$:
\begin{align}
&\ell_{X_1}=\ell^0_1+\Psi_1 \mathbf{e}_1^\tau \label{new65}\\
&\ell_{X_2}=\ell^0_2+\Psi_2 \mathbf{e}_2^\tau \label{new66}\\
&\ell_{X_3}=\ell^0_3+\Psi_3 \mathbf{e}_2^\tau +\Psi_4^{A_\tau}(C^1_x x_0 +C^2_x \ell_{X_1}+ C^3_x \ell_{X_2}) \label{new67}\\
&\mathbf{e}_1^\tau:=\begin{bmatrix} e_{-\tau}^{\tau\top},\cdots,e_{-\tau+N_c-1}^{\tau\top} \end{bmatrix}^\top\nonumber\\
&\mathbf{e}_2^\tau:=\begin{bmatrix} e_{-\tau}^{\tau\top},\cdots,e_{-1}^{\tau\top} \end{bmatrix}^\top \nonumber\\
&\Psi:=\begin{bmatrix} I &\!\!\! 0 &\!\!\! \cdots &\!\!\! 0 \\ 
A &\!\!\! I &\!\!\! \cdots &\!\!\! 0 \\ 
\vdots & \vdots & \ddots & \vdots \\ 
A^{\textit{N}-1} &\!\!\! A^{\textit{N}-2} &\!\!\! \cdots &\!\!\! I \end{bmatrix} \nonumber
\end{align}
where $\Psi_1$, $\Psi_2$, and $\Psi_3$ are the partitions of $\Psi$ as shown in Fig.~\ref{fig4}. By substituting (\ref{new65})--(\ref{new67}) and $d$ into (\ref{new53}), we can obtain $\ell_Y$. Then, $f_c$, $\ell_{f_{\bar{c}}}$, and $\ell_{Y}^{N_c}$ can be computed by substituting $\ell_Y$ into (\ref{new56}) and (\ref{renew62}). Finally, $\bar{f}_c$ is derived from (\ref{new60}) with computed $f_c$ and $\ell_{f_{\bar{c}}}$.

The resultant reduced optimal control problem for low-complexity mp-QP is given by
\begin{subequations}\label{renew70}
\begin{align}
\underset {\mathbf{U}_{c,\epsilon}}{\mathrm{min}}\quad &\textbf{U}_{c,\epsilon}^\top \bar{H} \textbf{U}_{c,\epsilon} +2\textbf{U}_{c,\epsilon}^\top \bar{f} \label{renew70a}\\
\mathrm{s.t.}\quad & x_{0}^e=\hat{x}^e(k),\; d=\hat{d}(k),\; u_{-1}=u(k-1) \label{renew70b}\\
& (\ref{rev53}),(\ref{rev54}),(\ref{rev55}),(\ref{rev30h}) \label{renew70C}
\end{align}
\end{subequations}
Then, we derive the explicit solution map as an optimizer function $\mathbf{U}_{c,\epsilon}^*(p)$ with respect to the parameter $p$ in (\ref{rev56}) by solving mp-QP of (\ref{renew70}). MPT3 Toolbox \cite{19} is used to solve the mp-QP. As a result, the parameter space is divided into several critical regions associated with each set of active constraints. Let $i\in\{ 1,\cdots ,n_{cr} \} $ denote the critical region index. The critical region from a set of active constraints $\mathcal{A}_i$ with an index $i$ can be described as a polyhedron with $\mathcal{H}$-representation:
\begin{align}
CR_{\mathcal{A}_i}=\lbrace p \;|\; H_i p \leq\mathbf{1} \rbrace.\label{eq44}
\end{align}
Then, when the current parameter is at a critical region with index $i$ (i.e., $p \in CR_{\mathcal{A}_{i}}$), the optimal solution $\mathbf{U}_{c,\epsilon}^*$ is described as a piecewise affine function of the parameter:
\begin{align}
\mathbf{U}_{c,\epsilon}^*(p)=F_i^* p+g_i^*.\label{eq45}
\end{align}
$F_i^*$ and $g_i^*$ are assigned for each critical region.

\section{Idle speed control using low-complexity EMPC}\label{sec5}

In this section, we demonstrate the ISC performance of the low-complexity offset-free explicit model predictive controller derived by the low-complexity mp-QP scheme developed in Section~\ref{sec4}. We utilized the high-fidelity engine model developed in Section~\ref{sec2} as the virtual engine, and the explicit model predictive controllers are implemented on the virtual ISC system. The characteristics of each implemented explicit controller is shown in Table~\ref{table1}.

\begin{table}[h]
\begin{center}
\caption{Detailed information of the designed low-complexity explicit controllers.}
\begin{tabular}{c | c c c}
\hline\rule{0in}{2.5ex} & EMPC1 & EMPC2 & EMPC3 \\
\hline\rule{0in}{2.5ex} Prediction horizon & 15 & 15 & 15  \\
\rule{0in}{2.5ex} Constraint horizon & 1 & 2 & 3 \\
\rule{0in}{2.5ex} \# of parameters & 5 & 8 & 11 \\
\rule{0in}{2.5ex} \# of critical regions & 39 & 1,211 & 28,300 \\
\hline
\end{tabular}
\label{table1}
\end{center}
\end{table}

The detailed implementation of low-complexity offset-free EMPC in ISC system is described in \textbf{Algorithm 1}.

\begin{table}[h]
\begin{center}
\begin{tabular}{l}
\hline
\textbf{Algorithm 1}. EMPC implementation in ISC system \rule{0ex}{2.5ex}\\
\hline\rule{0in}{2.5ex}\small\textbf{Initialize} $x^e(0)$, $d(0)$, $u^{\zeta}(0)$, $u^{w}(0)$ \\
\small $k \leftarrow 0$ (Beginning of ISC)\\
\small Apply $u^{\zeta}(0)$, $u^{w}(0)$ to the engine\\
\small Measure engine speed $y_m(0)$ from the sensor\\
\small Estimate $\hat{x}^e(1)$, $\hat{d}(1)$ by (\ref{rev29})\\
\small Update $x^e_0 \leftarrow \hat{x}^e(1)$, $d\leftarrow \hat{d}(1)$ \\
\small Update $u_{-1}^\zeta \leftarrow u^\zeta(0)$, $u_{-1}^w \leftarrow u^{w}(0)$\\ 
\small\textbf{Repeat} $k \leftarrow k+1$  \\
\small $\begin{matrix}  &  \end{matrix}$  Compute $p=[\bar{f}^\top_c,u_{-1}^{\zeta\top},u_{-1}^{w\top},\ell_{Y}^{N_c\top}]^\top$\\
\small $\begin{matrix}  &  \end{matrix}$  Derive $i_0$ s.t. $p\in CR_{A_{i_0}}$ via critical region search\\
\small $\begin{matrix}  &  \end{matrix}$  Compute $\textbf{U}^*_{c,\epsilon}$ by (\ref{eq45}) with $F_{i_0}^*$, $g_{i_0}^*$\\
\small $\begin{matrix}  &  \end{matrix}$  Obtain $u^{\zeta}(k)=u^{\zeta*}_0$, $u^{w}(k)=u^{w*}_0$ from $\textbf{U}^*_{c,\epsilon}$\\
\small $\begin{matrix}  &  \end{matrix}$  Wait for the next sampling instant \\
\small $\begin{matrix}  &  \end{matrix}$  (Beginning of the next sampling instant)\\
\small $\begin{matrix}  &  \end{matrix}$  Apply $u^{\zeta}(k)$, $u^{w}(k)$ to the engine \\
\small $\begin{matrix}  &  \end{matrix}$  Measure engine speed $y_m(k)$ from the sensor\\
\small $\begin{matrix}  &  \end{matrix}$  Estimate $\hat{x}^e(k+1)$, $\hat{d}(k+1)$ by (\ref{rev29})\\
\small $\begin{matrix}  &  \end{matrix}$  Update $x^e_0 \leftarrow \hat{x}^e(k+1)$, $d\leftarrow \hat{d}(k+1)$ \\
\small $\begin{matrix}  &  \end{matrix}$  Update $u_{-1}^\zeta \leftarrow u^\zeta(k)$, $u_{-1}^w \leftarrow u^{w}(k)$\\ 
\small\textbf{Until} the end of ISC \\
\hline
\end{tabular}
\end{center}
\end{table}

\begin{figure}[h!]
\begin{center}
\includegraphics[width=9cm]{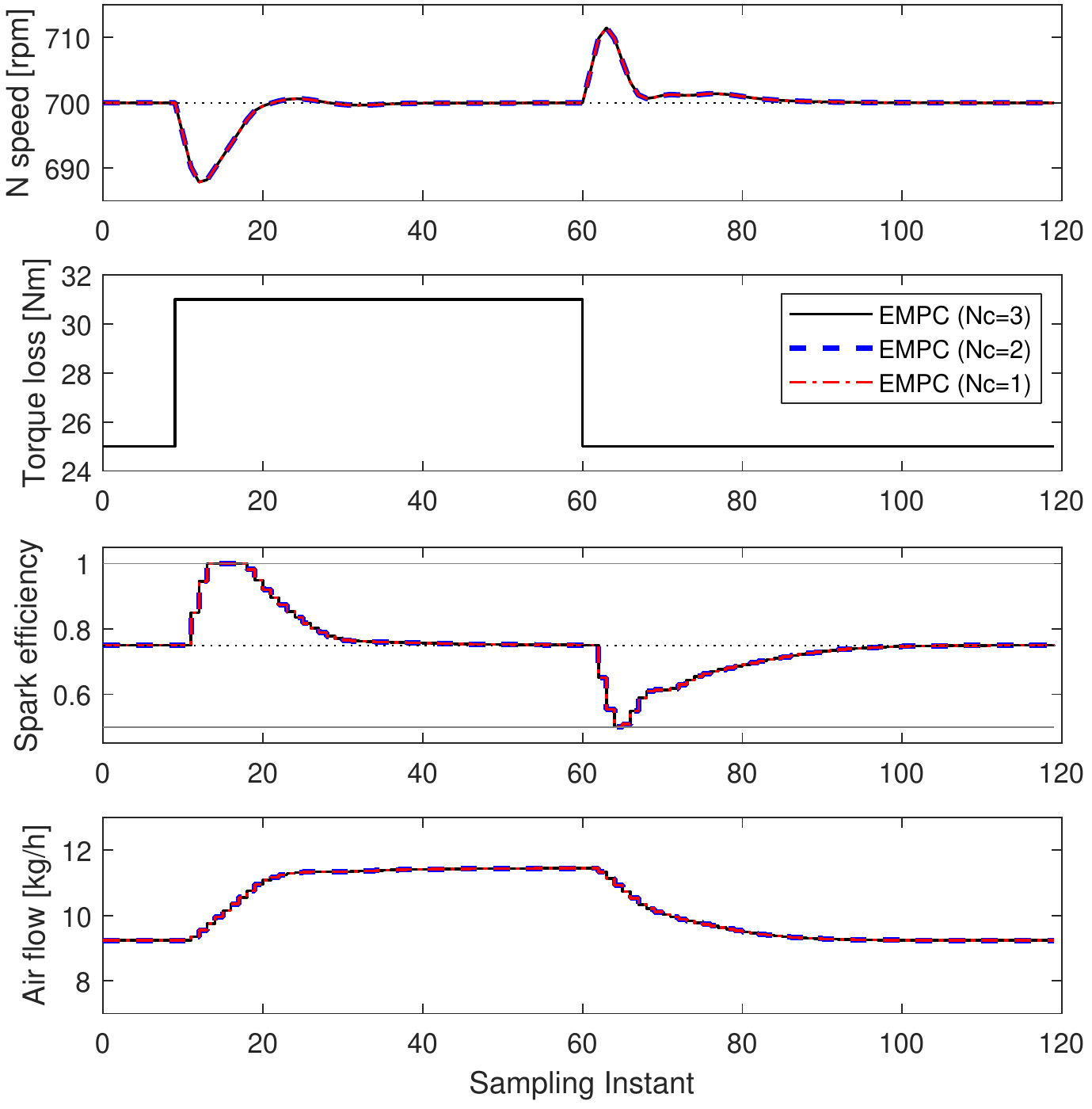}
\caption{Closed-loop trajectories of explicit idle speed controllers under step disturbance injection.}
\label{fig5}
\end{center}
\end{figure}

We first demonstrated the step disturbance rejection performance of the constructed  ISC system. Fig.~\ref{fig5} shows the closed-loop simulation results of the low-complexity offset-free EMPC in Table~\ref{table1}. All the controllers show the effective disturbance rejection performance for the stepwise torque loss change. In the early stage of torque loss change around the 15$^{th}$ sampling instant, the controllers actively exploit the spark efficiency which immediately affects the engine speed, and then gradually utilize the cylinder air flow which has intake center to torque center delay to affect the engine speed. Additionally, over the 20--40$^{th}$ sampling instants, the spark efficiency gradually returns to 0.75 for torque reserve, while the cylinder air flow gradually increases to the new steady-state value to reject the influence of the injected torque loss.

\begin{figure}[t!]
\begin{center}
\includegraphics[width=9cm]{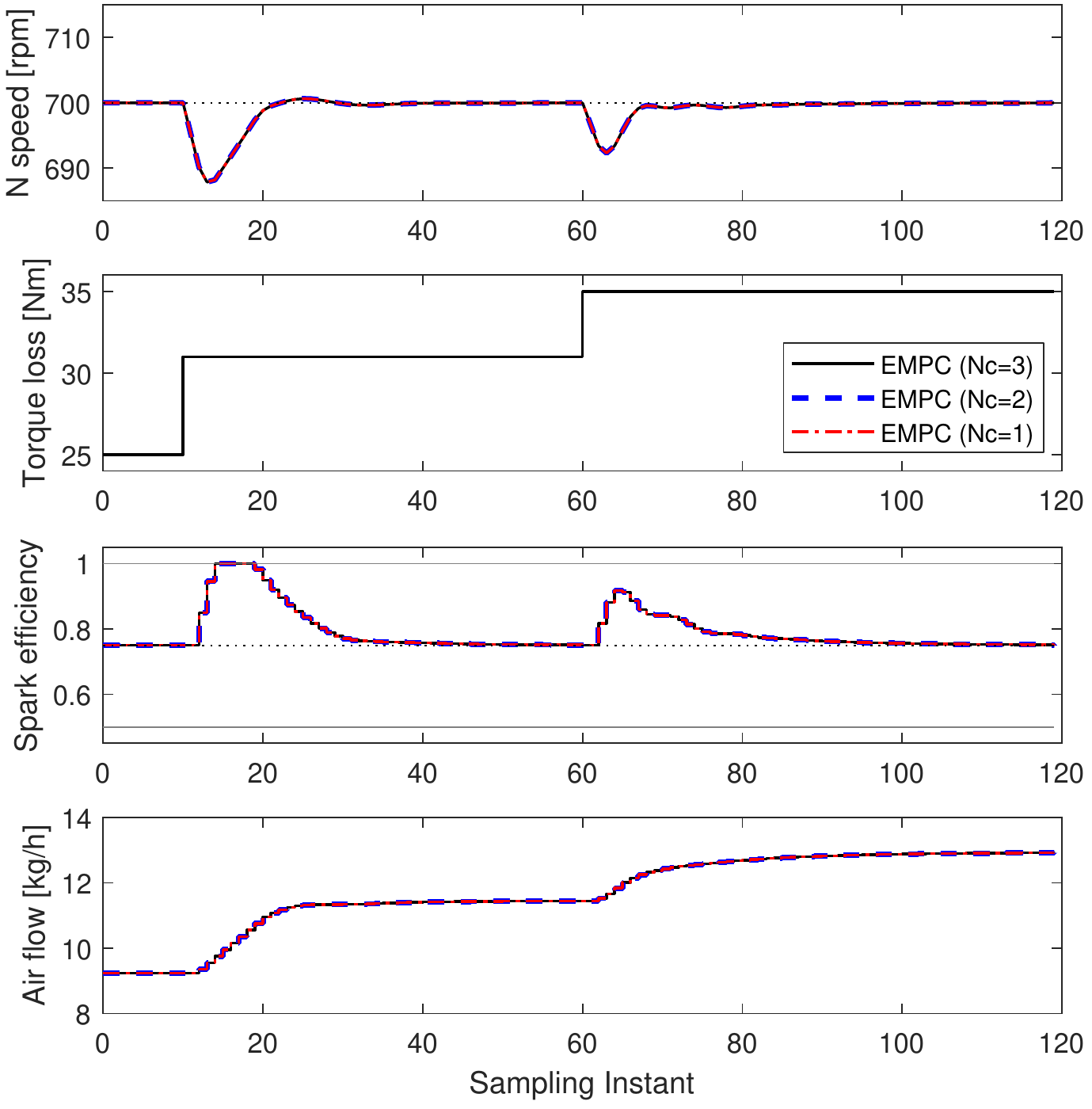}
\caption{Closed-loop trajectories of explicit idle speed controllers under step disturbance injection with torque reserve by spark efficiency degradation.}
\label{fig6}
\end{center}
\end{figure}

\begin{figure}[h]
\begin{center}
\includegraphics[width=9cm]{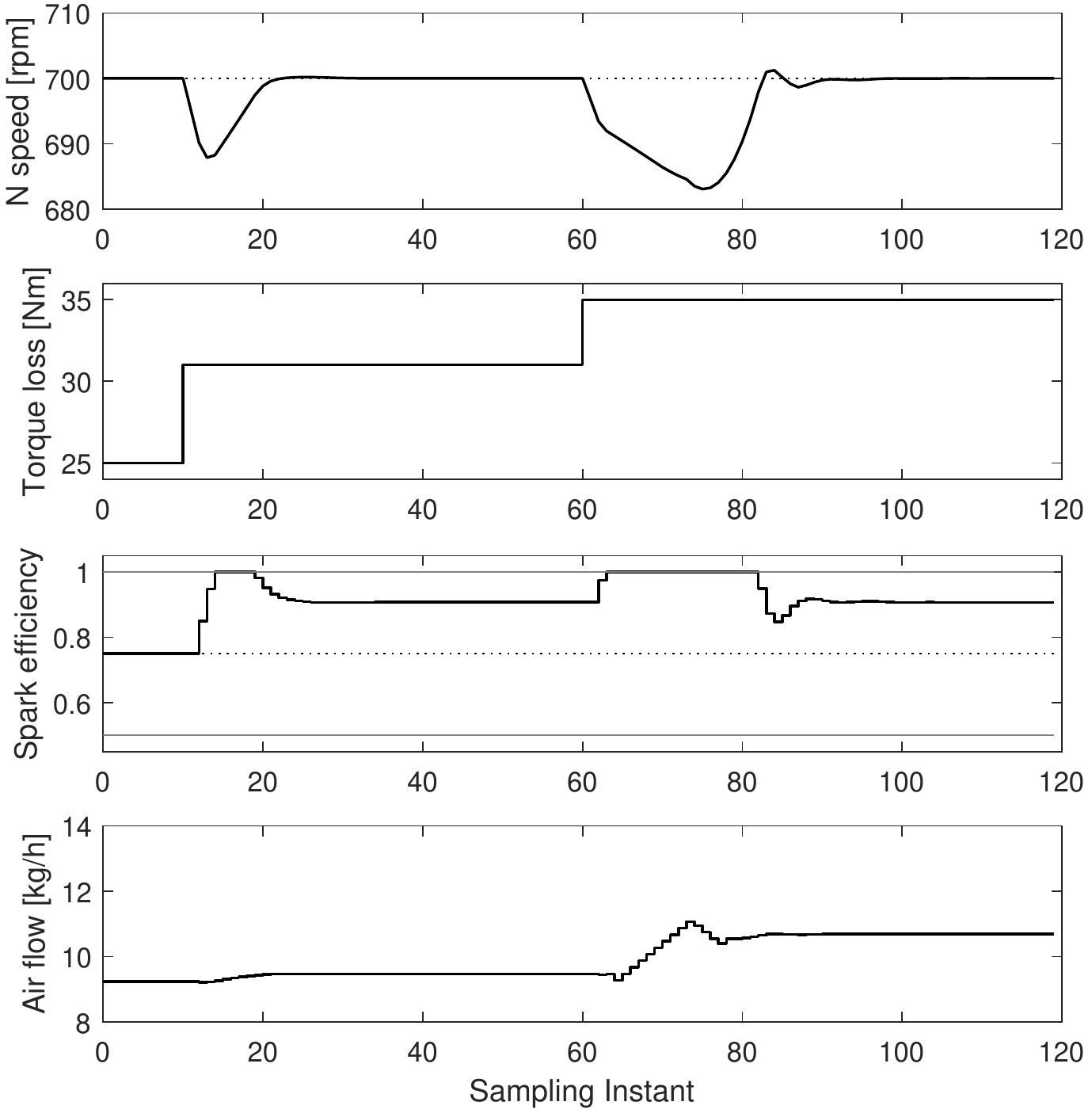}
\caption{Closed-loop trajectory of idle speed controller under step disturbance injection without torque reserve.}
\label{fig7}
\end{center}
\end{figure}

\begin{figure}[t!]
\begin{center}
\includegraphics[width=9cm]{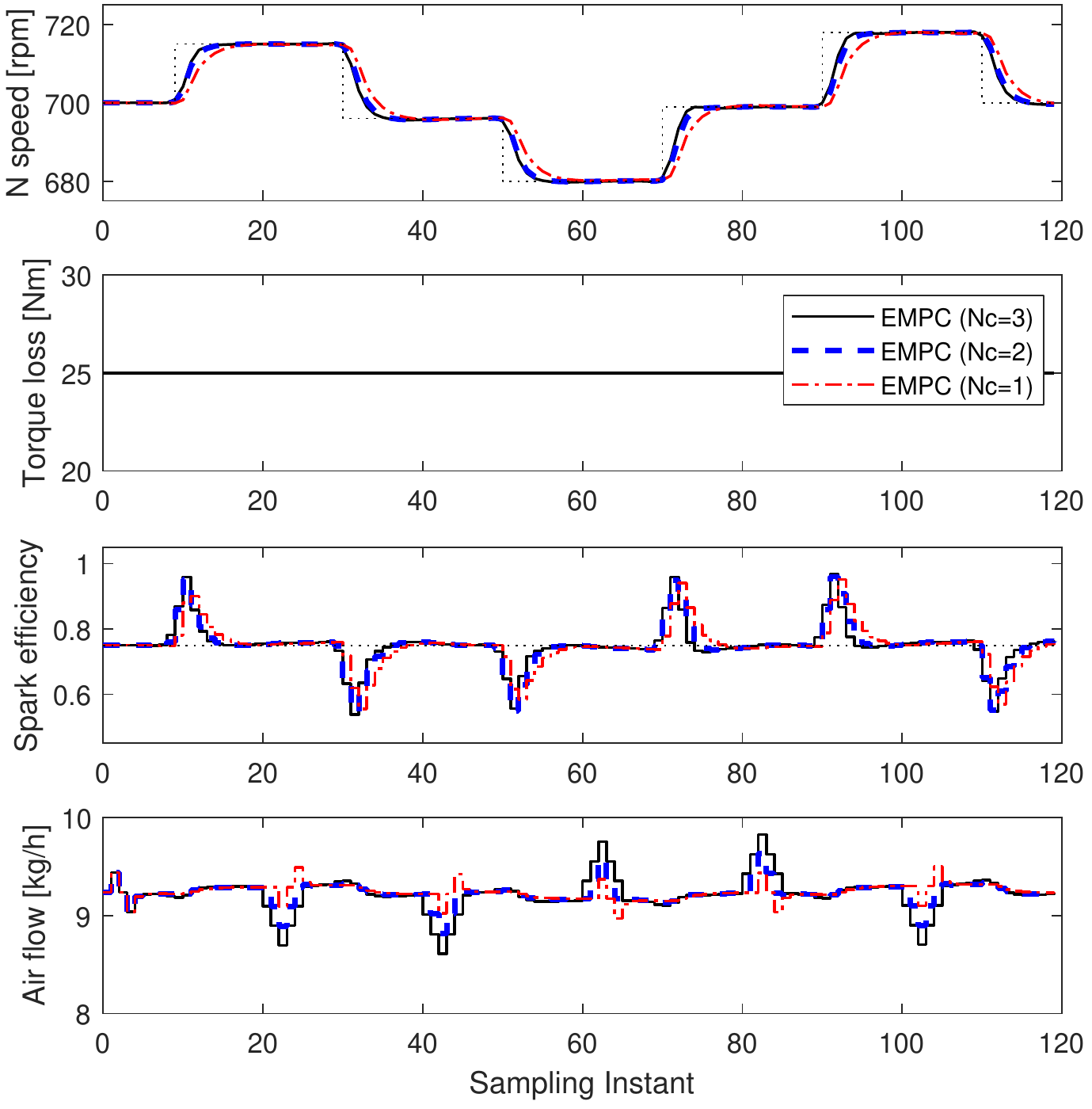}
\caption{Closed-loop trajectories of explicit idle speed controllers under set-point changes.}
\label{fig8}
\end{center}
\end{figure}

Since the ISC system controls one output, i.e., engine speed, by manipulating two inputs, i.e., spark efficiency and cylinder air flow rate, there are many combinations of steady-state values for manipulated variables that can drive the controlled variable to the desired set-point. This degree of freedom enables torque reserve via spark efficiency degradation so that the ISC system can immediately respond to torque loss. To demonstrate the effectiveness of the torque reserve, we compare the closed-loop performance of two cases where one case includes the torque reserve scheme with spark efficiency set-point, but the other case does not have torque reserve scheme. Fig.~\ref{fig6} shows the closed-loop simulation result from the control system with proposed low-complexity offset-free EMPC controllers which include the torque reserve scheme. The results show that the controllers can reject the additional step torque loss at the 60$^{th}$ sampling instant by utilizing the reserved spark efficiency over the 30--60$^{th}$ sampling instants. On the other hand, Fig.~\ref{fig7} shows the closed-loop simulation result from the control system without the torque reserve scheme. As we can see, the spark efficiency is not reserved during the 30--60$^{th}$ sampling instants, thus, the controller cannot immediately reject the additional step torque loss applied at the 60$^{th}$ sampling instant due to the limitation in available spark efficiency over the 60--80$^{th}$ sampling instants, and the engine speed trajectory shows considerable deviation from the set-point.

\begin{figure}[t!]
\begin{center}
\includegraphics[width=9cm]{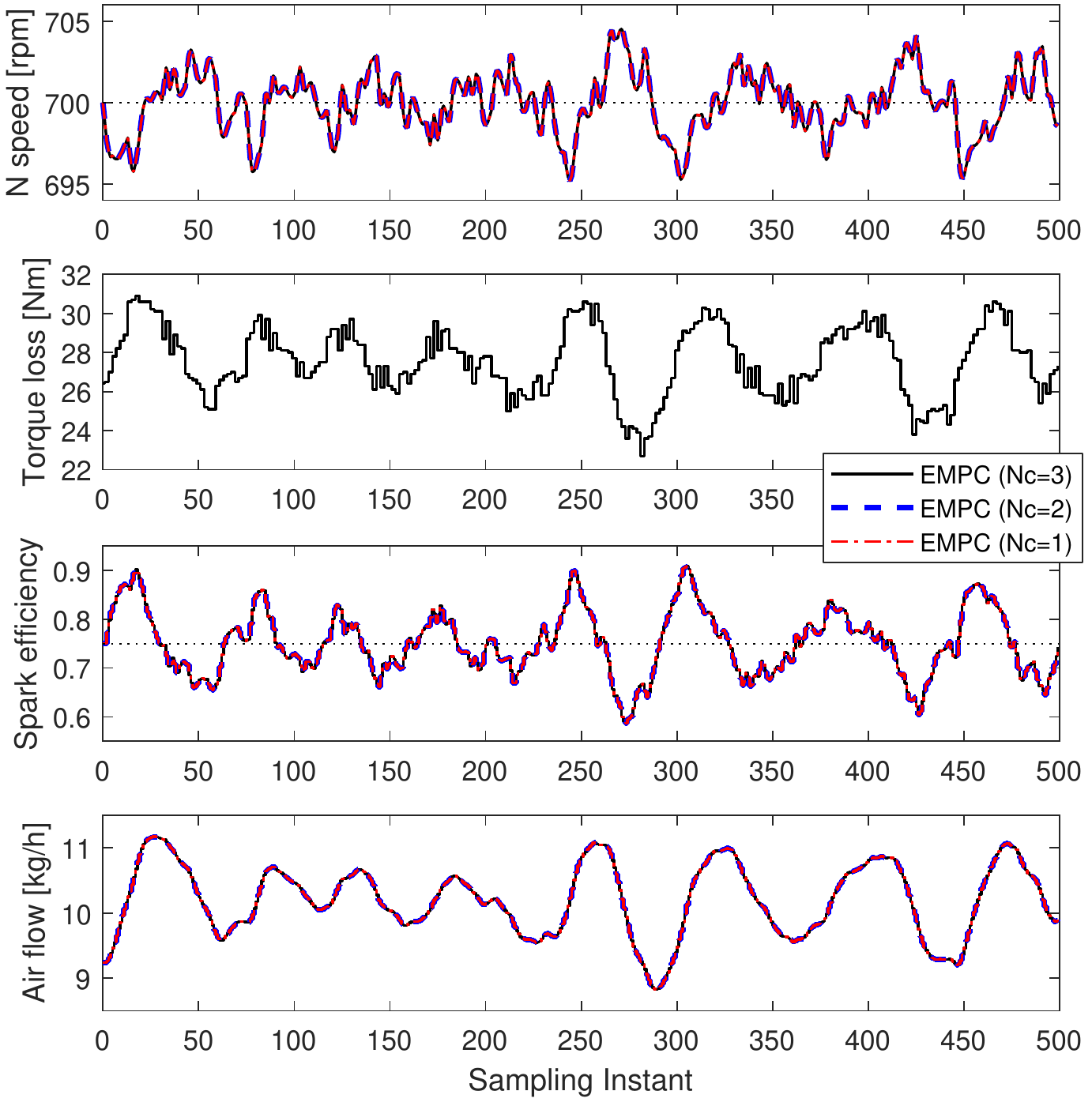}
\caption{Closed-loop trajectories of explicit idle speed controllers under the injection of torque loss data from a test vehicle.}
\label{fig9}
\end{center}
\end{figure}

\begin{figure}[t!]
\begin{center}
\includegraphics[width=9cm]{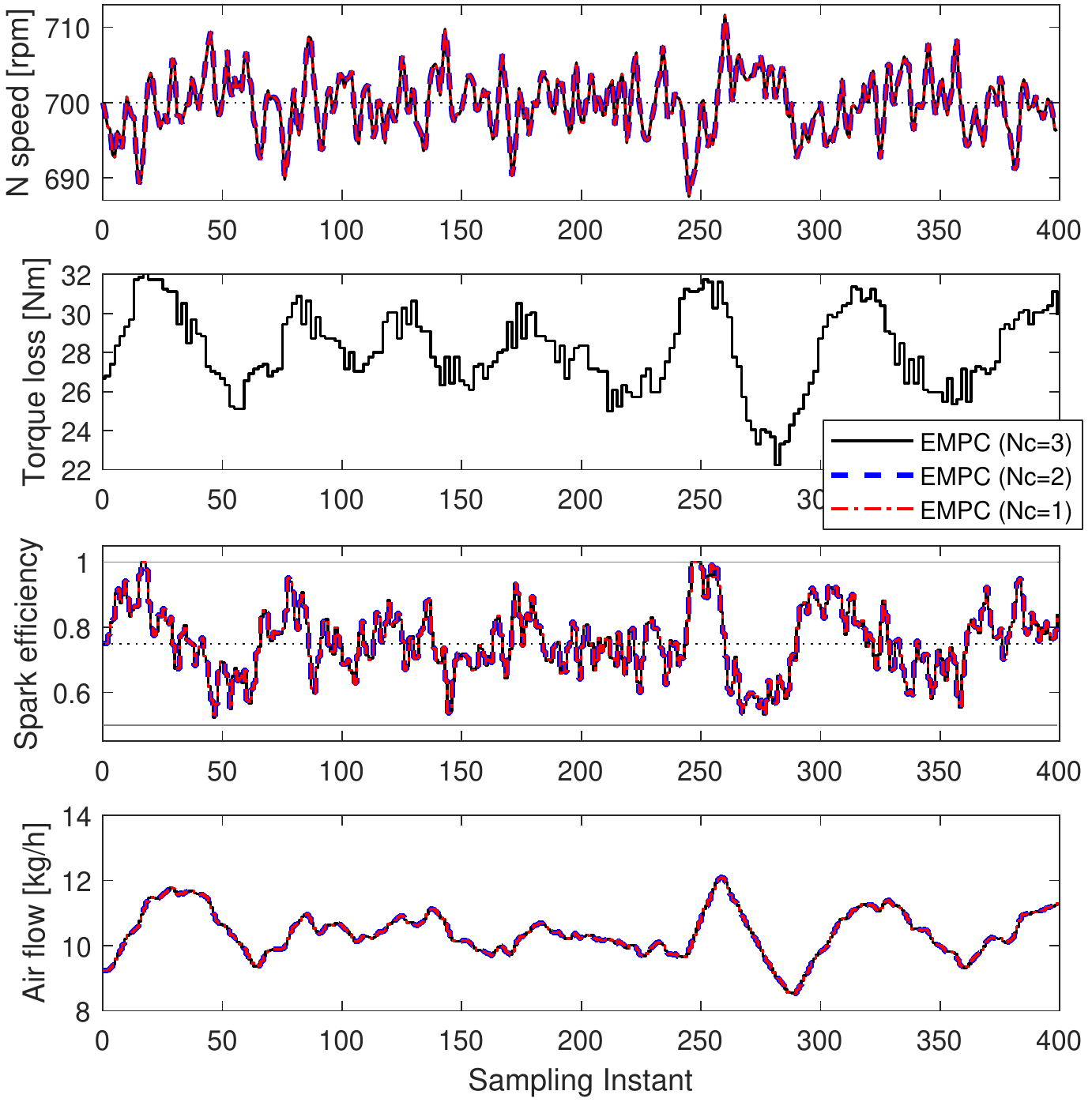}
\caption{Closed-loop trajectories of explicit idle speed controllers under real torque loss and measurement noise injection.}
\label{fig10}
\end{center}
\end{figure}

Fig.~\ref{fig8} shows the closed-loop trajectories from each low-complexity offset-free EMPC controller with the set-point change for engine speed under constant torque loss. All the controllers accomplish zero-offset tracking, but we can see a little difference in tracking performance and trajectories of the manipulated variables. The controllers with constraint horizons of 2 and 3 show better performance than that with constraint horizon of 1. Contrary to the result in Fig.~\ref{fig8}, the closed-loop trajectories in Figs.~\ref{fig5}~and~\ref{fig6} do not show much difference. This is because in the case of Figs.~\ref{fig5}~and~\ref{fig6}, since the model predictive controller cannot consider the future disturbance injection, the closed-loop result is dominantly affected by the performance of disturbance estimator which is identically applied to each controller. On the other hand, in the case of Fig.~\ref{fig8}, since the model predictive controllers can consider the set-point change within the prediction horizon, the closed-loop result is dominantly affected by the optimality of the controller.

\begin{figure}[h]
\begin{center}
\includegraphics[width=9cm]{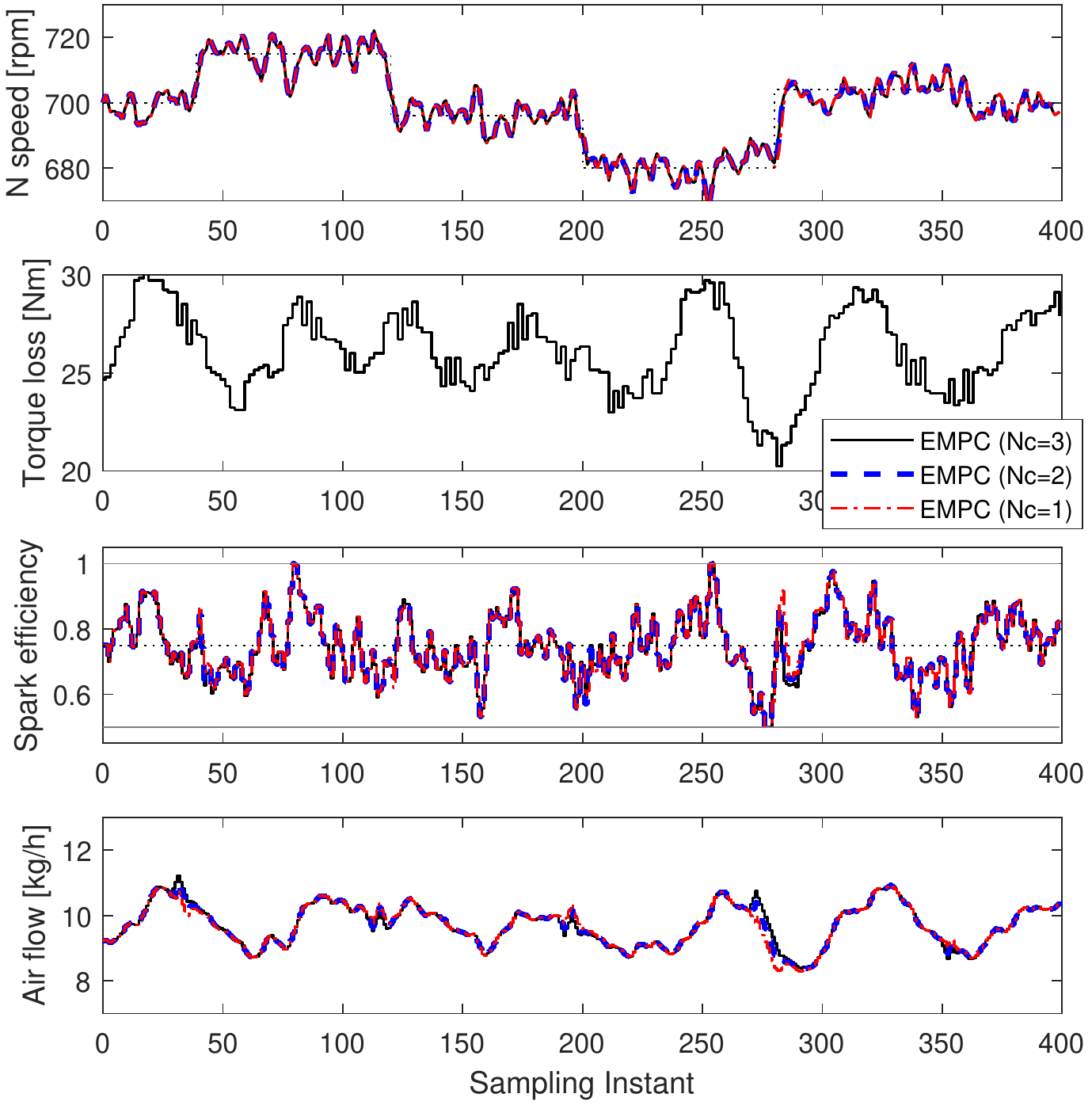}
\caption{Closed-loop trajectories of explicit idle speed controllers under real torque loss, measurement noise injection, and set-point changes.}
\label{fig11}
\end{center}
\end{figure}

Fig.~\ref{fig9} shows the closed-loop result of the developed idle speed controllers with the injection of torque loss data from a test vehicle. All the low-complexity offset-free EMPC controllers show proper disturbance rejection performance. The spark efficiency responds immediately to the injected torque loss while showing a similar trend to torque loss around the set-point 0.75, whereas the cylinder air flow responds afterward. This trend is because the controller intends to respond to the disturbance promptly by using the reserved spark efficiency which can influence the engine speed without intake to torque production delay, and to gradually utilize the cylinder air flow to reserve the spark efficiency.

We additionally applied the measurement noise with variance of 4 to the ISC system and demonstrated the closed-loop performance of the developed controllers in Fig.~\ref{fig10}. Though the resultant closed-loop trajectories of engine speed oscillate between 685 to 715 rpm due to the injected torque loss and measurement noise, the implemented controllers properly regulate the engine speed near the set-point 700 rpm. Then, a more challenging condition is applied to the ISC system via the set-point change for engine speed over time under the existence of torque loss and measurement noise. The closed-loop simulation result in Fig.~\ref{fig11} shows that all the implemented explicit model predictive controllers properly accomplish the tracking for the changed set-point for engine speed while rejecting the influence of injected torque loss under the noisy circumstance.

\section{Conclusion}\label{sec6}

We developed the ISC system for a 4-stroke SI-GDI engine to regulate the idle speed by rejecting the influence of torque loss via low-complexity offset-free EMPC in presence of system delay. The engine model was developed based on the first-principles, and then the parameter estimation was performed based on the data from a test vehicle. The control-oriented model was derived by linearizing and discretizing the engine model, and the effect of past state and input was augmented with the model to deal with the system delay. Then, the offset-free MPC system is designed to reject the influence of the torque loss while regulating the idle speed. Since the capacity assigned for the ECU is limited, EMPC scheme is also introduced to move the computational effort for on-line optimization to off-line. Additionally, a low-complexity mp-QP with constraint horizon is formulated and applied to reduce the complexity of solution map of the designed offset-free EMPC system.
 
The closed-loop simulation results showed that the developed ISC system properly regulated the idle speed in presence of torque loss, system delay, and measurement noise. In conclusion, the proposed low-complexity offset-free explicit model predictive ISC framework is expected to be an effective alternative as it is required to continually improve the performance of ISC to satisfy the increasingly stringent regulation on emission and fuel economy.

\section*{Acknowledgments}
This research was respectfully supported by Hyundai Motor Company. %We also thank the anonymous reviewers for their helpful comments on the original version of this paper.

% \cite{hartwich2008}
%Initial unit number change :\\

%\bibliographystyle{model2-names}
\bibliographystyle{elsarticle-num}
\bibliography{references}

%% \subsection*{A. Formulation of objective function in P$_0$ and P$_\ell^*$}

\end{document}